\documentclass[doublespacing]{elsart}
\usepackage[authoryear]{natbib}
\usepackage{epsfig}
\usepackage{amsmath}
\newcounter{sectioncounter}
\newcounter{Subsectioncounter}

\newcommand{\Appendix}[1]
{ {
\vspace{20pt}
\begin{center}
{\bf Appendix}
\end{center}
} }



\renewcommand{\thefootnote}{\fnsymbol{footnote}}

\begin{document}
\begin{frontmatter}

\title{Global disease spread: statistics and estimation of arrival times}

\author[Orsay,CNRS]{Aur\'elien Gautreau \corauthref{cor}},
\corauth[cor]{Corresponding author.}
\author[Orsay,CNRS,Turin]{Alain Barrat}
\author[CEA]{Marc Barth\'elemy}
\address[Orsay]{Univ Paris-Sud, 91405 Orsay,  France}
\address[CNRS]{CNRS, UMR 8627, 91405 Orsay, France}
\address[Turin]{Complex Networks Lagrange Laboratory, ISI Foundation,
Turin, Italy}
\address[CEA]{CEA-Centre d'Etudes de
Bruy{\`e}res-le-Ch{\^a}tel, D\'epartement de Physique Th\'eorique et
Appliqu\'ee BP12, 91680 Bruy\`eres-Le-Ch\^atel, France}

\begin{abstract}
  We study metapopulation models for the spread of epidemics in which
  different subpopulations (cities) are connected by fluxes of
  individuals (travelers). This framework allows to describe the
  spread of a disease on a large scale and we focus here on the
  computation of the arrival time of a disease as a function of the
  properties of the seed of the epidemics and of the characteristics
  of the network connecting the various subpopulations. Using
  analytical and numerical arguments, we introduce an easily
  computable quantity which approximates this average arrival time. We
  show on the example of a disease spread on the world-wide airport
  network that this quantity predicts with a good accuracy the order
  of arrival of the disease in the various subpopulations in each
  realization of epidemic scenario, and not only for an average over
  realizations. Finally, this quantity might be useful in the
  identification of the dominant paths of the disease spread.
\end{abstract}


\end{frontmatter}
%\vspace*{0.25cm}

%\begin{multicols}{2}

\section{Introduction}

In our modern world, the existence of various transportation means has
strongly affected the way in which infectious diseases spread among
humans. In fact, it has become unavoidable to take into account in
the study of the geographical spread of epidemics the various
long-range heterogeneous connections typical of modern transportation
networks, This naturally gives rise to a very complicated evolution of
epidemics characterized by heterogeneous outbreaks patterns
\citep{cohen2000,cliff2004,Colizza:2006a,Colizza:2006b}, as recently
documented at the world-wide level in the SARS case
(http://www.who.int/csr/sars/en).
%~\citep{SARS}.

Such complex phenomenon can be tackled at different granularity levels,
including very detailed agent-based simulations, complicated social and
spatial structures, complex contact networks,
etc.~\citep{May:1992,Hethcote:1984,Morris:1996,Keeling:1999,Pastor:2001a,Pastor:2001b,May:2001,Ferguson:2003,Meyers:2005,Chowell:2003,Eubank:2004}.
In particular, a very important class of models in modern epidemiology
describes the propagation between different interconnected subpopulations with
the help of the so-called metapopulation models. In these models, the disease
spreads inside each subpopulation (often assumed
homogeneously mixed) and is transmitted between different subpopulations by a
``coupling'' depending on the model. In the case of human infectious diseases,
this coupling is caused by the movements of individuals and depends on the
transportation means relevant at the spatial scale chosen for the description
of the spread.

In the context of large scale spread, air-transportation represents a major
channel of epidemic propagation, as pointed out in the modeling approach to
global epidemic diffusion of Rvachev and
Longini~\citep{Longini:1985,Baroyan:1969} and by similar studies on the
behavior of specific outbreaks such as pandemic influenza, HIV or
SARS~\citep{Longini:1988,Grais:2003,Brownstein:2006,Flahault:1991,Hufnagel:2004,Colizza:2006a,Colizza:2006b,Colizza:2007a}.
In this case, the relevant metapopulation model considers as subpopulations
the inhabitants of the various world cities and the air travel of infectious
individuals results in the spread of the disease from one city to another.

Taking into account the global airline transportation infrastructure
in metapopulation models has recently become possible thanks to the
availability and analysis of large scale databases
\citep{Barrat:2004,Amaral:2004}, and to the always increasing computer
capacities. The word wide airport network (WAN) is described by a
complex weighted graph, in which the airports are the vertices and the
weighted links represent the presence of direct flight connections
among them (the weights corresponding to the number of available seats
on each connection). This network, which consists of more than $3000$
nodes and $18000$ links, displays small-world properties as well as
strongly heterogeneous topology and traffic properties. For many
different dynamical phenomena occurring on complex networks
\citep{barabasi02,Doro:2003,Pastorbook:2003,Boccaletti:2006}, the
presence of such emerging properties has been shown to imply the
breakdown of the standard results. This is particularly true for
epidemiology ~\citep{Pastor:2001a,Pastor:2001b,May:2001} where
classical results about epidemic thresholds do not apply if the
heterogeneity of the network is too large. This therefore calls for a
systematic investigation of the epidemic spread in the framework of
metapopulation models defined on complex networks such as the world
airport network. Such investigations can be carried out in various
parallel and complementary directions. On the one hand, the availability
of large scale datasets has recently allowed for the development of tools
for intensive computational epidemiology
\citep{Colizza:2006a,Colizza:2006b,Colizza:2007a}, which can be used
for example for scenario evaluations~\citep{Colizza:2007a} and risk
assessment. On the other hand, these computational tools can also be
used for the study of fundamental properties of these metapopulation
models, and to understand the main mechanisms of the disease spread
and the role of the various complex properties of the transport
network. In particular, \cite{Colizza:2006a,Colizza:2006b} have
investigated how the properties of the global air travel network
affect the heterogeneity and the predictability of spreading
patterns. Very recently moreover, the effect on the epidemic threshold
of both the network properties and the endogenous reaction was studied
using analytical and numerical tools typical of statistical
physics~\citep{Colizza:2007b,Colizza:2007c}.

In this paper, we tackle the problem of determining the arrival time
of an epidemics spread in a metapopulation model consisting in
subpopulations (cities) coupled by a transportation network. We stress
that this study is different from the ones of \cite{bart:2006} and
\cite{Barrat:2005}, in which each node of the network is an individual
(and not a full subpopulation), and where links represent the possibility of
the contamination between individuals. The main result of the present
paper is to exhibit an easily computable quantity depending only on
the networks characteristics and which approximates the average
arrival time. This quantity is only an approximation of the exact
arrival time but we show on the example of the world-wide disease
spread that it allows to rank the cities according to the arrival of
the disease.

The paper is organized as follows. We present in section 2
%\ref{sec:model} 
the Rvachev-Longini metapopulation model for the
epidemics spreading, and in 
section 3, we study the 
%\ref{sec:defs} 
arrival time of a disease for simple topologies such 
as a one-dimensional line of cities. We
then extend these results to the case of complex networks and we propose a
quantity which is a good approximation to the 
average arrival time in the various
nodes or subpopulations connected through an arbitrary network of connections.
We finally show in section 4,
%\ref{sec:num}, 
using numerical simulations of epidemic spreading on the worldwide
airport network, that our approximation is not only valid for the
average arrival time, but is also able to give the order of arrival of
the disease in the various cities with a good precision for each
spreading realization. A brief account of some of the results presented
here can be found in \citep{jstatlett}.

%-----------------------------------------------------------
\section{The Rvachev-Longini metapopulation model}
\label{sec:model}

The Rvachev-Longini model~\citep{Longini:1985} was initially introduced to
describe the spread of the 1968-69 Hong-Kong flu at the worldwide level. This
metapopulation model uses two different levels of description: the various
subpopulations are given by the inhabitants of the $58$ large cities
corresponding to the $58$ largest airports, and homogeneous mixing is assumed
at the individual city level. In this model, the propagation of the disease
from one subpopulation to another is due to individuals traveling on the
air-transportation network between these airports~\citep{Longini:1985}.

The evolution of the number $I_i$ of infectious individuals in each city $i$ can
thus be written as the sum of two terms
\begin{equation}
\partial_tI_i=K(\{X_i\})+\Omega(\{I_j\}) \ ,
\end{equation}
where the first term $K$ of the rhs describes the (epidemic) reaction
process inside each subpopulation (city), due to the interaction of
individuals in the various possible states $X$ ($X=S, L, I, R...$
depending on the population compartimentalization into Susceptible,
Latent, Infected, Recovered... individuals). The second term of the
rhs represents the incoming and outgoing fluxes of infectious
individuals to and from other cities $j$. This model therefore
considers a simplified mechanistic approach with a markovian
assumption in which individuals are not labeled according to their
original subpopulation, and where at each time step the same traveling
probability applies to all individuals in the subpopulation, without
any memory of their previous
locations \citep{Longini:1985,Hufnagel:2004,Colizza:2006a,Colizza:2006b}.
The travel term $\Omega$ depends therefore on the air-transportation network: if
the weight $w_{ij}$ represents the number of passengers traveling from
$i$ to $j$ per unit of time ($w_{ij}=0$ if there is no direct
connection between $i$ and $j$), and $N_i$ the population of the city
$i$, it is reasonable to assume that the probability per unit time
that an individual in city $i$ travels to city $j$ is given by
$w_{ij}/N_i$.  In the case of a simple SI model, the time evolution of
the number $I_i$ of infectious infividuals in city $i$ is therefore
given by:
\begin{equation}
\partial_tI_i=\lambda I_i\left(t\right)\frac{N_i-I_i\left(t\right)}{N_i}
+\sum_j\frac{w_{ji}}{N_j}I_j-\sum_j\frac{w_{ij}}{N_i}I_i \ ,
\label{RLcont}
\end{equation}
where $\lambda$ is the transmission rate. Similar equations can be written for
the other compartments in the various possible models (SIS, SIR...), by
modifying the first term $K$ of the rhs in Eq.~(\ref{RLcont}) accordingly.

This original formulation has an important drawback. Indeed, it is
deterministic, since it considers only expectation values, while both
epidemic spreading and travel of individuals are inherently stochastic
processes. The number of infectious individuals is then treated
as a continuous variable and, even if the initial condition (at $t=0$)
of the spreading consists in one single infectious individual in a
given city $i_0$, all cities have a non-zero density of infectious at
any time $t>0$.  In order to tackle the problem of the arrival time of
the spread in the various subpopulations, this pathology has to be
accounted for.  A first possibility, already presented by
\cite{Longini:1985}, to avoid this unrealistic situation consists in
considering that each travel term $w_{ij}I_i/N_i$ is present only when
$w_{ij}I_i/N_i > 1$, i.e. in putting a threshold term
$\theta(w_{ij}I_i/N_i-1)$, corresponding to the fact that the expected
number of travelers has to be larger than unity. In this approach
however the ``quantity'' of individuals traveling is still a
continuous variable.  Another possibility, which consists in treating
all quantities as integers and going back to the microscopic
stochastic processes, can be necessary for detailed modeling purposes
\citep{Colizza:2007a}. This solution is however computationally quite
demanding, so that we will consider in all numerical simulations
performed in this paper an intermediate framework: the travel term is
treated as in the stochastic generalization described by
\cite{Colizza:2006a,Colizza:2006b} (see also \cite{Hufnagel:2004}),
where the number of individuals traveling on each connection is an
integer variable randomly extracted at each time step. More precisely,
each of the $[I_i]$ (the integer part of $I_i$) infectious individuals
has a probability $p_{ij}=w_{ij}\Delta t/N_i$ to go from $i$ to $j$ in
the time interval $\Delta t$. The numbers $\xi_{ij}$ of infectious
individuals traveling on the various connections existing from $i$ to
other cities $j$ are given by a set of stochastic variables which
follow then a multinomial distribution \citep{Colizza:2006b}
\begin{equation}
P\left(\{\xi_{ij}\}\mid\{I_i\}\right)=
\frac{[I_i]!}{\left(
[I_i] - \sum_j\xi_{ij}\right)!\prod_j\xi_{ij}!}
\times\prod_j p_{ij}^{\xi_{ij}}\left(1-\sum_j p_{ij}\right)^{[I_i]-\sum_j\xi_{ij}}
\label{multinom}
\end{equation}
where the mean and variance of the stochastic variables are
$\langle\xi_{ij}\left([I_i]\right)\rangle = w_{ij}\Delta t [I_i]/N_i$
and $Var\left(\xi_{ij}\right)=\frac{w_{ij}}{N_i}\Delta t\left(1-
  \frac{w_{ij}}{N_i}\Delta t\right)[I_i]$.  On the other hand, the
endogenous growth (inside each city) will be treated for simplicity as
a deterministic evolution of densities of infectious and susceptible
individuals. We have moreover checked numerically that the inclusion
of stochastic effects described by noise terms in the evolution
equations \citep{Colizza:2006a,Colizza:2006b} do not change our
results.  Finally, we will present the results essentially for the
simple SI model, since arrival times will basically depend on the
first stage of the disease development. We have however carried out
numerical simulations as well in the SIS and SIR case, with consistent
results (see section 4).

It is interesting to note that in realistic cases such as a disease
propagation on the WAN, most weights are symmetric ($w_{ij}=w_{ji}$)
\citep{Barrat:2004}
but the probabilities of travel from one city to another are not
($p_{ij}\neq p_{ji}$). The travel therefore effectively occurs as a
random diffusion on a directed weighted complex network. Moreover, an
important difference with the case of random walks on complex
networks~\citep{Rieger:2003} comes from the endogenous evolution
inside each subpopulation. Indeed, as soon as an infectious individual
reaches a city, a contamination process starts and the number of
random infectious walkers is not constant. Such an intricate behavior
has also important consequences on the existence of epidemic
thresholds, as studied recently by~\cite{Colizza:2007b}.

As a last general remark, we note that the simple topological distance
between nodes should naturally play a role in the context of arrival
times, but does not contain all the information needed to characterize
such a process, nor does a priori the optimal weighted distance, which
takes into account the weights \citep{Wu} but not the populations nor
the endogenous epidemic evolution. Moreover, since most transportation
networks are small-world networks (even when taking into account
weights), many cities lie at the same topological distance from a
given seed, but will be reached at very different times.

%----------------------------------------------------
\section{Arrival times statistics}

\label{sec:defs}

In this section, we study the arrival time of the first infectious
individual in the various cities, starting with very simple topologies
for the transportation network.

\subsection{Two cities : arrival time distribution}

We start with the case of two connected cities ($0$ and $1$). We
denote the link weight by $w_{01}=w$, the population of the first city
by $N_0=N$, the number of infectious individuals in city $0$ by
$I_0=I$. The probability to jump from $0$ to $1$ during the time
interval $\Delta t$ is $p=\frac{w}{N}\Delta t$. The initial condition
is given by $I_0(t=0)=I^0$, $I_1(t=0)=0$, i.e., infectious individuals
are present only in city $0$. We consider that the travel events
occur as instantaneous jumps (of probability $p$ for each individual)
at discretized times.  The probability that the first infectious
individual arrives at time $t_1=t =n\Delta t$ in city $1$ is then
given by
\begin{equation}
P\left(t_1=n \Delta t\right)=
\left(1-\left(1-p\right)^{{[I_0(n\Delta t)]}}\right) \times
\prod_{i=1}^{n-1}\left(1-p\right)^{{[I_0(i\Delta t)]}} \ ,
\label{probaI}
\end{equation}
which expresses the fact that at least one successful ``jump'' from
$0$ to $1$ of an infectious individual occurs at time $n \Delta t$, and none 
at previous times. In Eq. (\ref{probaI}),
the quantity $[I_0]$ denotes the integer part of $I_0$. In
order to obtain the probability density of the arrival time in city $1$, 
we first note that in real-world systems the number
of travelers is usually small with respect to the total population of a city:
$p=w\Delta t/ N \ll 1$ (with any reasonable $\Delta t$). In this
limit, Eq.~(\ref{probaI}) becomes
\begin{equation}
P\left(t_1=t\right) =p [I_0(t)]
e^{-p\sum_{0< i <n}{[I_0\left(i\Delta t\right)]}} \ .
\end{equation}
We will also assume that the travel probability is sufficiently small so that
the number of infectious individuals $I_0(t)$ in the city $0$ grows
substantially before the city $1$ is contaminated. Writing $1 \ll I_0(t) \ll
N$ (the second inequality is due to the fact that in realistic cases only
small fractions of the population are contaminated), we thus consider the
following approximation: $[I](t) \approx I_0(t) \approx I^0e^{\lambda t}$ (as
long as $t < t_1$). Using as well the standard approximation $\Delta
t\sum_{0<i<n}[I_0\left(i\Delta t\right)]= \int_0^tI_0\left(\tau\right)d\tau$,
we obtain for $I^0=1$
\begin{equation}
P(t) dt=
\frac{w}{N}e^{\lambda t-\frac{w}{N\lambda}e^{\lambda t}} \Theta(t) dt \ ,
\label{distrfin}
\end{equation}
where $\Theta(t)$ is the Heavyside function which ensures the positivity
of the arrival time. 
The distribution (\ref{distrfin}) is a Gumbel distribution with average
$\langle t_1\rangle =
\frac{1}{\lambda}\left(\ln\left(\frac{N\lambda}{w}\right)-\gamma\right)$,
where $\gamma$ is the Euler constant. 
The contribution of the negative $t$ in
the distribution has to be negligible before $1$, which reads
$\int_{-\infty}^0P\left(t\right)dt=\frac{w}{N\lambda}\ll1$. This range of
validity corresponds also to the hypothesis of $\lambda\langle
t_1\rangle \gg 1$ (the time to increase $I_0$ by one individual
is small with respect to the time scale of the epidemic arrival in city $1$).

\begin{figure}[p]
\begin{center}
\includegraphics*[width=12.5cm]{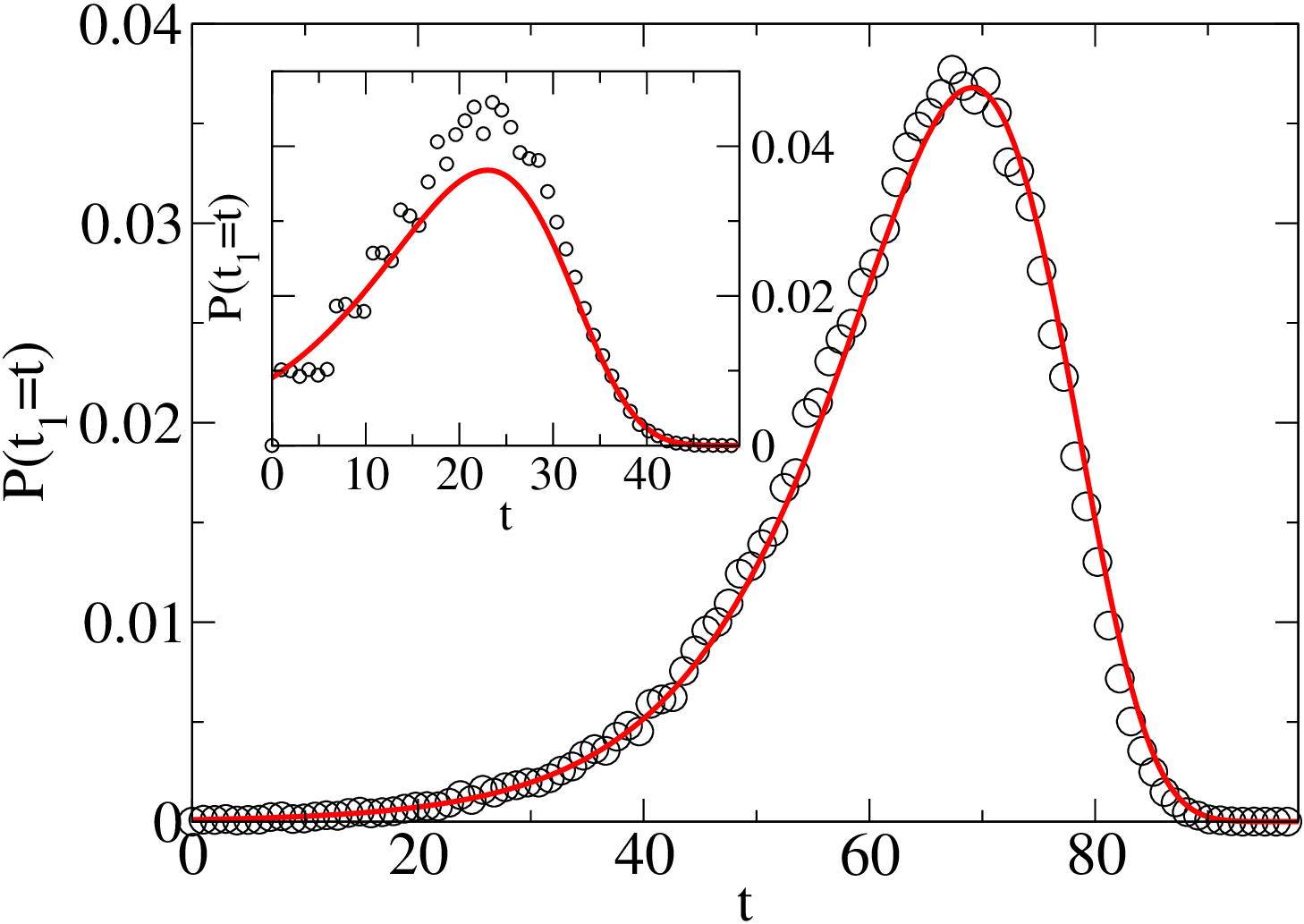}
\end{center}
\caption{ \small Two cities model, distribution of arrival times
  $t^{\left(1\right)}$ in numerical simulations with
  $\frac{w}{N\lambda}=10^{-2}$. \textit{Inset} : distribution of
  $t^{\left(1\right)}$ with $\frac{w}{N\lambda}=10^{-1}$. The
  continuous line is the analytical approximation given by
  Eq.~(\ref{distrfin}). Here and in the following, unless specified,
  $N=10^6$ and $\lambda=0.1$.  } \vspace*{.5cm}
\label{gumbel}
\end{figure}

Figure \ref{gumbel} displays the results of numerical simulations of
the propagation between two cities, using discrete stochastic travel events as 
described in section 2. The whole distribution of arrival
times in the second city shows a very good agreement with the form
(\ref{distrfin}), obtained through the continuous approximations
detailed above. When $w/(N\lambda)$ is not small enough, stronger
deviations are obtained; this is expected since the hypothesis $1 \ll
I_0(t)$ for $t < t_1$ is less valid. Nonetheless, the overall shape of
the distribution is still in good agreement (Inset of
Fig.~\ref{gumbel}).

\subsection{One-dimensional line}

We now consider a one-dimensional line of subpopulations $i\geq 0$,
with an initial condition given by one infectious individual in city
$0$. We denote by $w_i$ the travel flux between city $i$ and city
$i+1$, $t_i$ the arrival time of the first infectious in city $i$, and
$\Delta_i=t_i - t_{i-1}$ ($t_0=0$). The quantity $t_n$ is thus the sum
of the random variables $\Delta_i$ for $i=1,...,n$, and the average is
given by $\langle t_n \rangle = \sum_{i=1,...,n}\langle
\Delta_i\rangle $. These random variables $\Delta_i$ are a priori
correlated and not identically distributed which implies that the
central limit theorem can not be used in general.

\subsubsection*{Homogeneous line}

In the case of a homogeneous line with uniform populations and weights
($N_i=N$, $w_i=w$), one expects that the distributions of $\Delta_n$
becomes independent of $n$ at large $n$, with a well-defined average
$\lim_{n \to \infty} \langle \Delta_n\rangle=\langle \Delta\rangle$,
so that $t_n \approx n\langle \Delta\rangle$. The issue then is the
computation of $\langle \Delta\rangle$.
Two time scales only are present:
$N/w$ which represents the average time for an individual to travel
from one city to the next, and $1/\lambda$ which represents the
typical transmission time of the disease. 
Dimensional analysis thus implies that the
adimensional quantity $\lambda\langle\Delta\rangle$ has the form
\begin{equation}
\lambda\langle\Delta\rangle=F\left(\frac{w}{N\lambda}\right) \ .
\label{dim_an}
\end{equation}

In order to estimate the unknown function $F$, a first possibility is
to consider that at large $n$ the spread consists in an evolving
epidemic front obeying the continuous equation
$\partial_tI(x,t)=\lambda SI+w\partial^2_xI(x,t)$.
Looking for a traveling wave solution~\citep{Murray} leads to a front
speed $v=2\sqrt{\frac{\lambda w}{N}}$, and thus to
\begin{equation}
\langle\Delta\rangle_{front} \approx\frac{1}{2\lambda}
\sqrt{\frac{N\lambda}{w}} \ .
\label{deltafront}
\end{equation}

Another possibility consists in using the results of the previous
subsection, and assume that the average of $\Delta_n$ remains close to
the arrival time in the first city
$\langle\Delta_n\rangle\approx\langle\Delta_1\rangle=\langle
t_1\rangle$ which yields
\begin{equation}
\langle \Delta\rangle_{Gumbel} =\frac{1}{\lambda}\left[
\ln\left(\frac{N\lambda}{w}\right)-\gamma\right] \ .
\label{deltagumbel}
\end{equation}
This approximation neglects the fact that
$I_{n-1}(t)$ increases between $t_{n-1}$ and $t_n$ due both to the endogeneous
growth in city $n-1$ and to the arrival of infectious individuals from
city $n-2$, while for the computation of $\langle t_1\rangle$, only
the endogeneous growth of $I_0$ has to be considered. It can therefore
be expected that $\langle \Delta\rangle_{Gumbel}$ will overestimate
the real $\langle \Delta\rangle$.

Finally, we can also consider a deterministic
formulation, in which travel between $i$ and $i+1$ occurs only if
there are enough infectious individuals in $i$, i.e.  $I_i(t)$ has
reached a certain threshold $\theta$, and is then treated as
continuous. In this framework, each city is considered infected only
if the number of infectious individuals is above $\theta$, and $t_n$ is
defined as the first time that $I_n$ reaches the threshold:
$I_n(t_n)=\theta$. Between $t_{n-1}$ and $t_{n}$, no travel can
therefore occur out of $n$, and we can write $\partial_tI_n=\lambda
I_n+\frac{w}{N}I_{n-1}$. During $[t_{n-1},t_{n}]$, under the
realistic hypothesis that $I_n\ll N_n$, and that $I_{n-1}$ is in a
first approximation given by
$I_{n-1}=e^{\lambda\left(t-t_{n-1}\right)}$, we obtain the equation
$1=\frac{w}{N}\Delta_ne^{\lambda\Delta_n}$. The difference between the
arrival times in two successive cities is therefore given by
\begin{equation}
\langle \Delta\rangle_{det}=\frac{1}{\lambda}W\left(\frac{N\lambda}{w}\right)
\label{deltadet}
\end{equation}
where $W$ is known as the Lambert W function.

\begin{figure}[p]
\begin{center}
\includegraphics*[width=12.5cm]{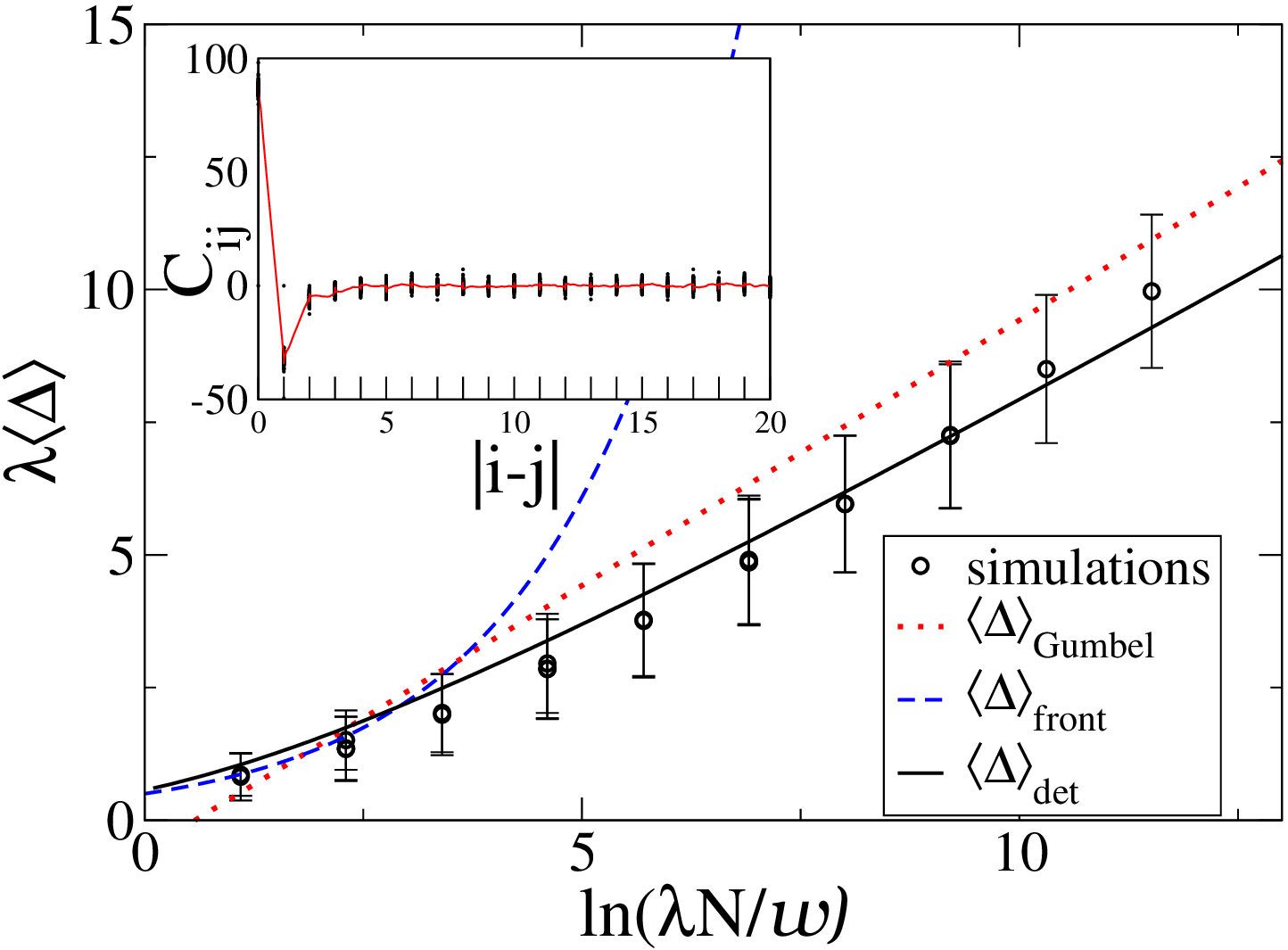}
\end{center}
\caption{ \small Slope $\langle \Delta\rangle$ of $\langle t_n\rangle$
vs $n$ on a line of length $500$.  The average $\langle t_n\rangle$ is
taken over $500$ realizations of the spreading, and the slope is
measured at large $n$ (only the cities with $n \ge 100$ are
considered). The error bars represent the variance of
$\Delta$. $\lambda \in [10^{-2},1]$ and $w \in [10,10^5]$,
$N=10^6$. Full line: $\langle \Delta\rangle_{det}$.  Dotted line:
$\langle\Delta\rangle_{Gumbel}$.  Dashed line:
$\langle\Delta\rangle_{front}$. Inset: Correlations
$C_{ij}=\langle \Delta_i\Delta_j\rangle -\langle \Delta_i\rangle
\langle \Delta_j\rangle $ versus $\mid i-j\mid$ on a line of length
$500$, the average is done over $500$ realizations.  }\vspace*{0.5cm}
\label{grandn}
\end{figure}

In order to test these various analytical approaches, we consider
numerical simulations of the stochastic model described in section 2,
for identical cities of population $N=10^6$ located on a
one-dimensional line, with uniform travel fluxes $w$ between
successive cities. The measured average arrival times are, as
expected, proportional to $n$ at large $n$ (not shown).  Figure
\ref{grandn} displays the corresponding slope as $\lambda \langle
\Delta\rangle$ versus $\ln(\lambda N/w)$. Various values of $N$,
$\lambda$ and $w$ with the same ratio $\lambda N/w$ yield the same
$\lambda \langle \Delta\rangle$, as predicted from the theoretical
dimensional analysis (\ref{dim_an}).  As shown in Fig. \ref{grandn},
$\langle \Delta \rangle_{front}$ is in agreement with the simulations
only at small $\lambda N/w$: the slowest the travel, the less a
spatial continuous approximation is valid. On the other hand, both
$\langle\Delta\rangle_{det}$ and $\langle\Delta\rangle_{Gumbel}$
display reasonable agreement, and in particular correctly capture the
increase with $\ln(\lambda N/w)$ for large $\lambda N/w$. As expected,
$\langle\Delta\rangle_{Gumbel}$ slightly overestimates
$\langle\Delta\rangle$.  We also show in the inset of
Fig. \ref{grandn} the correlations between the $\Delta_n$'s. These
correlations vanish rapidly with the distance along the line. Negative
correlations can however be observed between $\Delta_{n-1}$ and
$\Delta_{n}$. Such phenomenon can be understood as follows. Assume
that, in a given spreading realization, $\Delta_{n-1}$ is small. In
this case, $I_{n-2}\left(t_{n-1}\right)$ will be unusually small
too. The ``reservoir'' of infectiousness defined by city $n-2$ will
thus transmit less infectious individuals to $n-1$ at times $t >
t_{n-1}$, and therefore the subsequent contamination of city $n$ will
be slower, leading to a larger $\Delta_n$.

\begin{figure}[p]
\begin{center}
\includegraphics*[width=12.5cm]{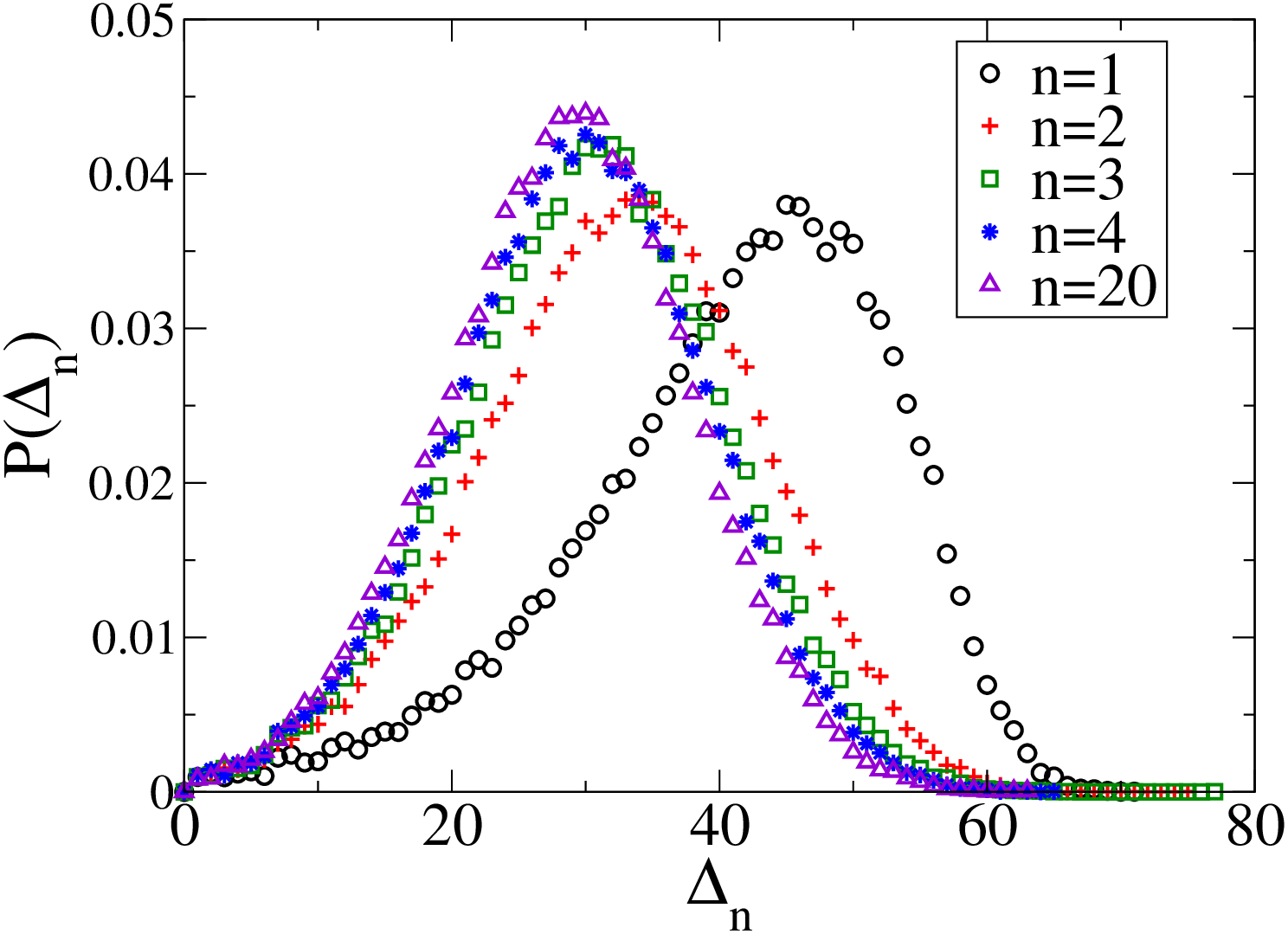}
\end{center}
\caption{ \small Distribution of $\Delta_n=t_n-t_{n-1}$ for various values of
  $n$. For small $n$, the distribution is close to a Gumbel and for large $n$,
  it evolves to a Gaussian. $w/(N\lambda)=10^{-2}$,  $N=10^6$,
    $\lambda=0.1$.}\vspace*{0.5cm}
\label{gaussian}
\end{figure}

\begin{figure}[p]
\begin{center}
\includegraphics*[width=12.5cm]{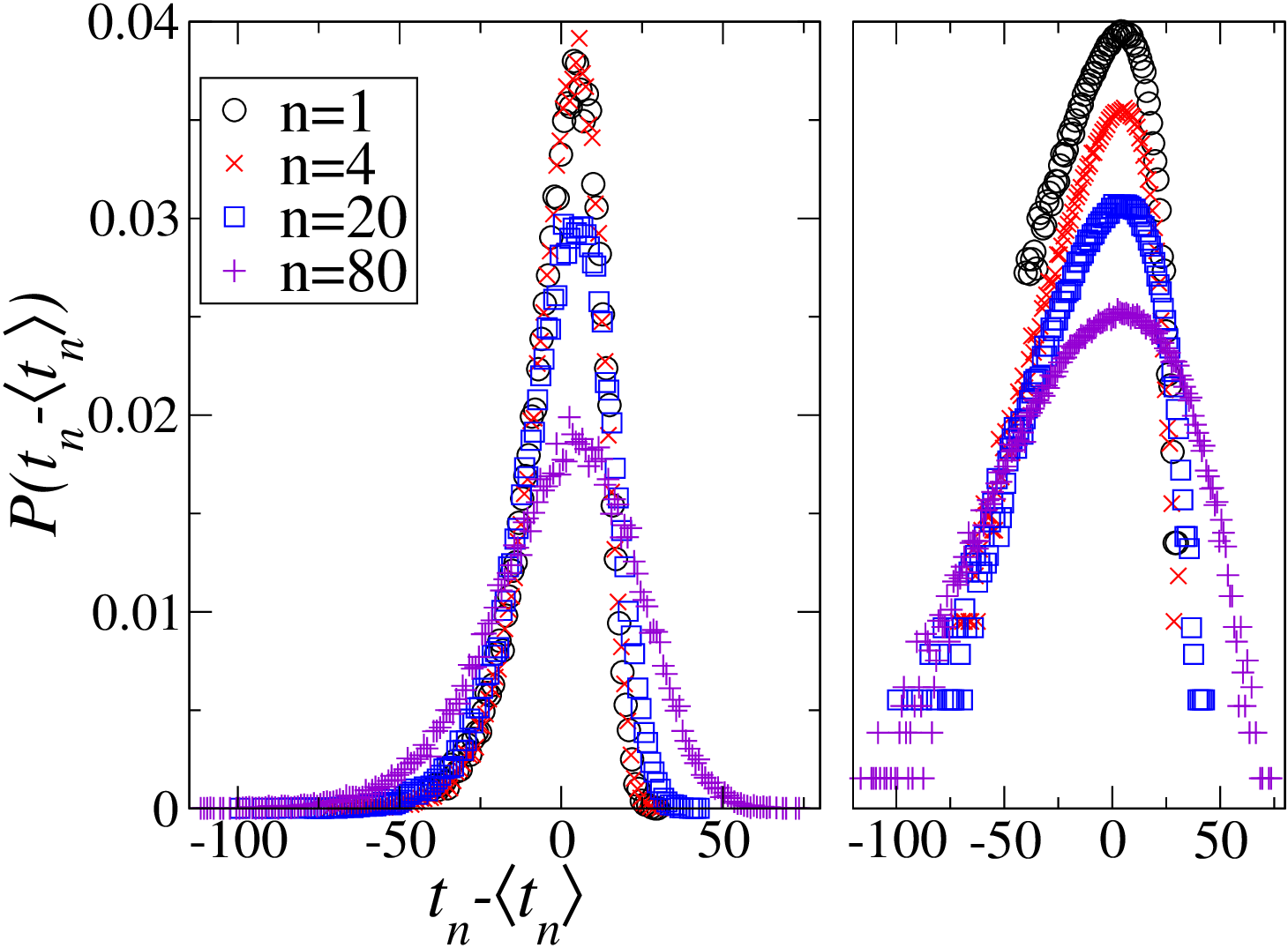}
\end{center}
\caption{ \small Distribution of $t_n$ for various values of $n$. The
  distribution remains close to Gumbel shapes even at large $n$. The
  plot on the right is in log-scale on the y-axis and the curves are
  shifted vertically for clarity. $w/(N\lambda)=10^{-2}$, $N=10^6$,
  $\lambda=0.1$.}\vspace*{0.5cm}
\label{distrtime}
\end{figure}

The numerical simulations allow also to measure the whole
distributions of arrival times and of their intervals. In
Fig.~\ref{gaussian}, we show $P(\Delta_n)$ for different values of
$n$. The Gumbel shape is only valid for $n=1$, and for large $n$ more
symmetric (Gaussian-like) distributions are obtained. On the other
hand, Fig. \ref{distrtime} shows that the distribution of the arrival
times themselves display asymmetric Gumbel-like shapes even at large
$n$.

\subsubsection*{General populations and weights}

\begin{figure}[p]
\begin{center}
\includegraphics*[width=12.5cm]{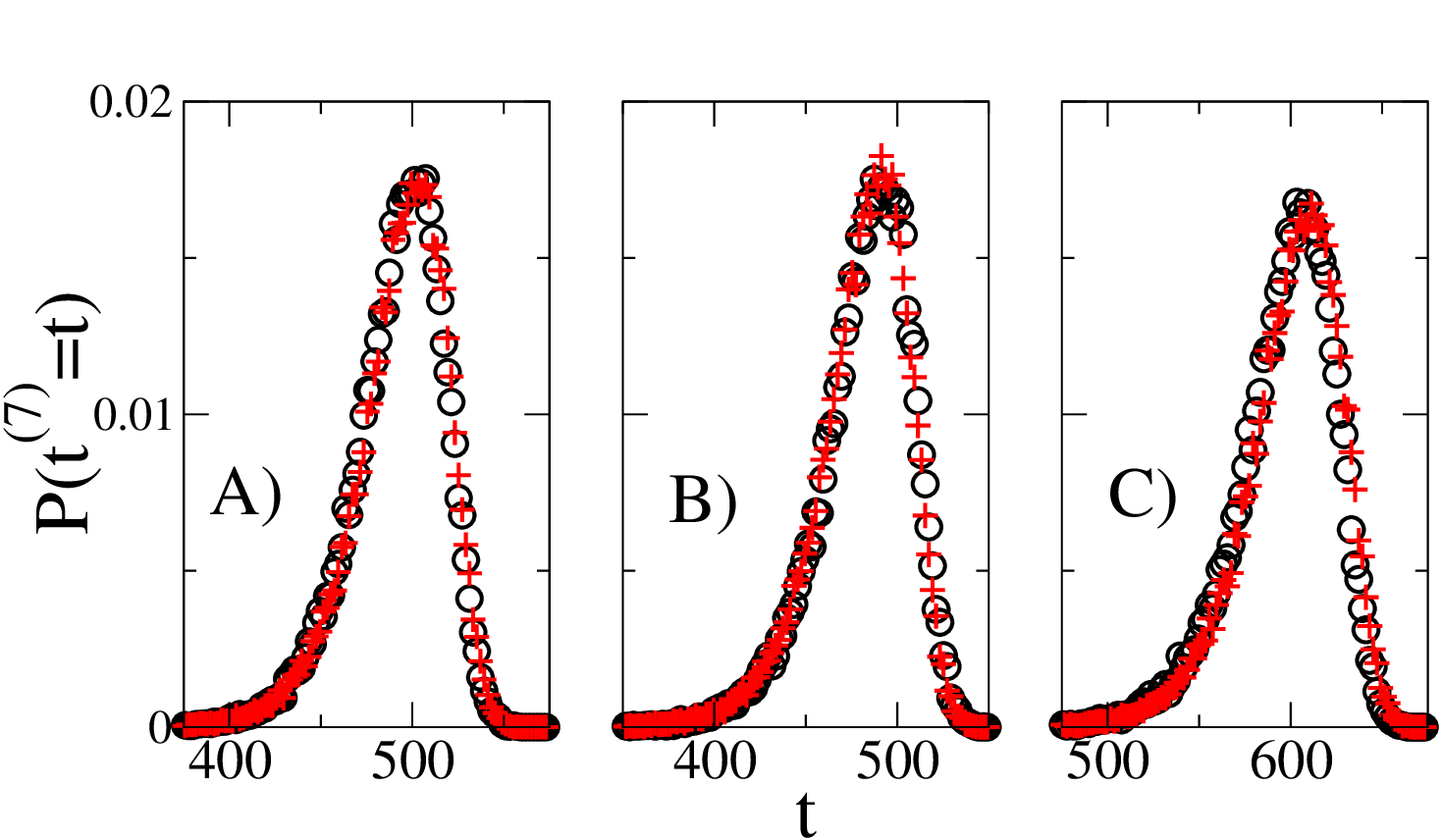}
\end{center}
\caption{ \small (A-C) Black circles indicate the arrival time
  distribution on a line at the city $\#$7 obtained for a fixed random set of
  populations $\{N_i\}$ and weights $\{w_i\}$ ($3$ different
sets are used in the $3$ graphs). Red crosses correspond to the
  same distribution obtained A) with uniform travel $w_i=\bar{w}$, and
  populations $\{N_i\}$; B) with uniform populations $N_i=\bar{N}$, and weights
  $\{w_i\}$; C) with uniform populations $N_i=\bar{N}$ and uniform weights
  $w_i=\bar{w}$.  }\vspace*{0.5cm}
\label{sym}
\end{figure}

We now consider the general case of heterogeneous populations $N_i$ and fluxes
$w_i$. The application of the same considerations as in the case of a
homogeneous line leads to $\langle t_n \rangle = \sum_{i=1}^n \langle
\Delta_i \rangle$.  Since our main goal is to understand the case of complex
networks which usually have the small-world property (i.e. a small diameter
varying typically as $\log N$), we will now focus on the properties of $t_n$
at small $n$. Assuming that the average of $\Delta_i$ remains close to the
formula for the arrival time in the first city ($\langle \Delta_i \rangle
\approx \langle t_1(\lambda,w_i,N_i)\rangle$), we obtain
\begin{equation}
\lambda \langle t_n \rangle \approx
\chi_n \equiv
\ln\left[\prod_{i=0}^{n-1}\frac{N_i\lambda e^{-\gamma}}{w_i}\right] \ .
\label{averaget}
\end{equation}
This expression links the expected arrival time in city $n$ of a
stochastic spreading event to the quantity $\chi_n$ which depends on
the properties of the line on which the spreading propagates. It is
symmetric by permutations of weights and populations, and Figure
\ref{sym} shows in fact that the whole distribution of arrival times,
obtained through numerical simulations of the discrete stochastic
dynamics, and not only its average, respects this symmetry. In the
numerical simulations, heterogeneous populations and travel fluxes are
uniformly distributed ($w_i \in [10,2000]$ and $N_i \in
[10^5,2.10^7]$). Figure \ref{sym} shows that the distribution of
arrival times is invariant when one replaces (i) all the random
weights by their geometrical mean $\overline{w}=(\prod_{i=0\dots
n}w_i)^{1/n}$; (ii) all the random populations by their geometrical
mean $\overline{N}=(\prod_{i=0\dots n}N_i)^{1/n}$; (iii) all weights
by $\overline{w}$ and all populations by $\overline{N}$.  The ratios
of the average arrival times for these different sets $\langle
t(\{w_i\},\{N_i\}) \rangle/\langle t(\overline{w},\{N_i\})\rangle$, $\langle
t(\{w_i\},\{N_i\}) \rangle/\langle t(\{w_i\},\overline{N})\rangle$ and
$\langle t(\{w_i\},\{N_i\})\rangle/\langle t(\bar{w},\bar{N})\rangle$ stay
very close to $1$, with deviations at most of the order of $5\%$.

\begin{figure}[p]
\begin{center}
\includegraphics*[width=12.5cm]{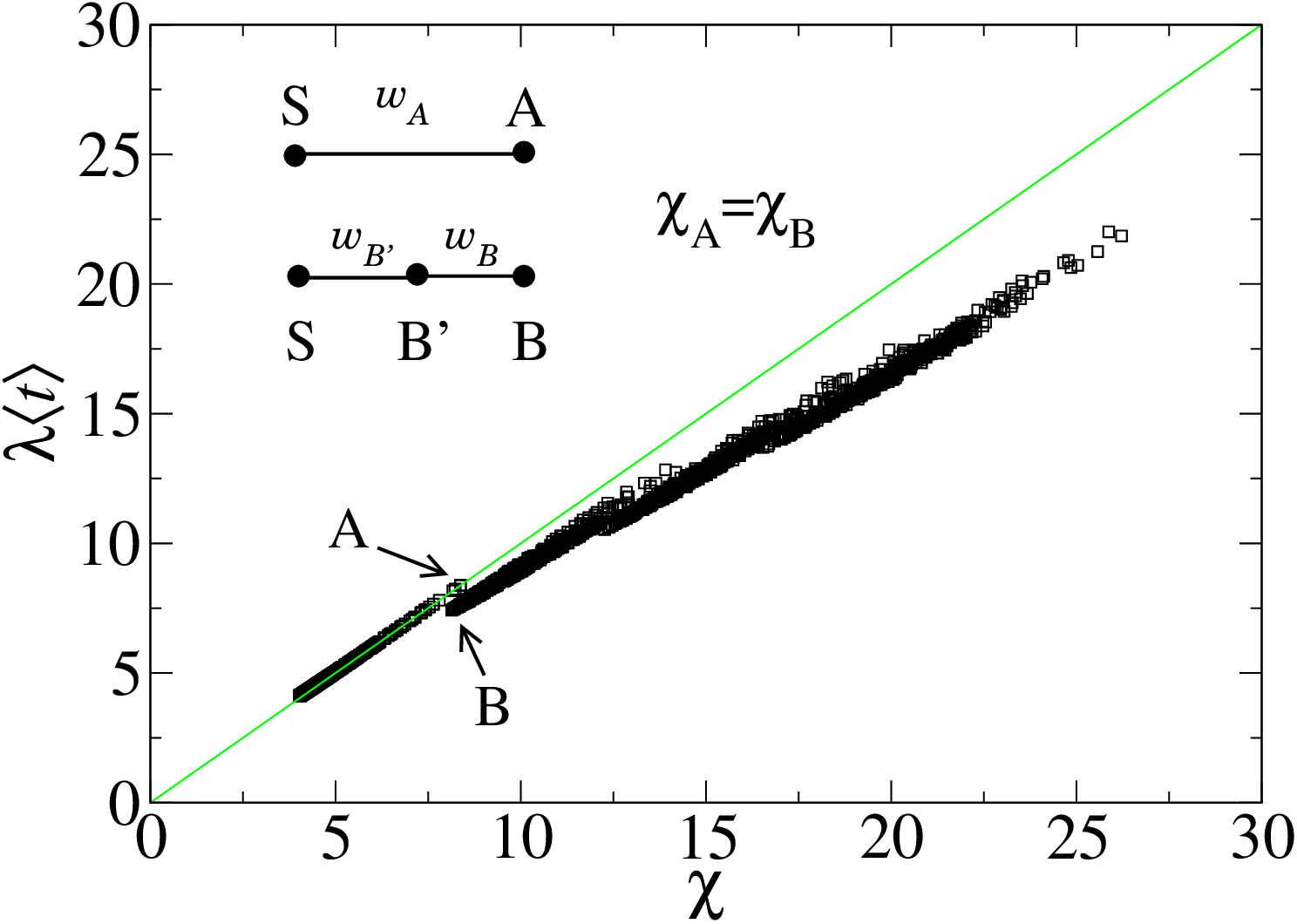}
\end{center}
\caption{ \small $\lambda\langle t\rangle $ vs $\chi$ for $5$ cities
   connected on a line.  Each point correspond to the average arrival
   time in one of the city of the line, for one realization of the
   weights $w$, averaged over $10^3$ realizations of the epidemic
   spread. Results are presented for $100$ different random sets $\{w_i\}$.
   } \vspace*{.5cm}
\label{chi}
\end{figure}

We test in Fig.~\ref{chi} the relation (\ref{averaget}) which predicts the 
average arrival time measured in numerical simulations of the stochastic model.
We use for simplicity uniform populations $N=10^6$, and
we measure, for each fixed set of weights, the arrival time in each city, 
averaged over $10^3$ stochastic realizations of the spread. Figure~\ref{chi}
displays $\lambda\langle t_i\rangle$ versus $\chi_i$ for the various
cities of a $5$ cities line, for $100$ different random sets of weights
(each point corresponds to one given set of weights). We first note 
that $\lambda\langle t_i\rangle$ is a monotonous function of
$\chi_i$. The figure also clearly shows that $\chi$ systematically
overestimates the average arrival time. This is due, as already  
explained in the discussion of Eq. (\ref{deltagumbel}), to the fact that
the use of $\langle \Delta\rangle_{Gumbel}$ neglects correlations, and
in particular the travel between cities $n-2$ and $n-1$ for the
estimate of the arrival time in $n$. Since this travel term increases
$I_{n-1}$, it increases the probability that an infectious travels from
$n-1$ to $n$, and therefore reduces $t_n$. This overestimation effect
is not present for the first group of points that lie on the diagonal
of the figure and correspond to cities directly connected to the
seed. For this first group of cities, the average arrival time is indeed
correctly given by $\chi$. On the other hand, the small overestimation
can already be observed for cities at distance $2$ of the seed.  For
example, points $A$ and $B$ highlighted in Fig. \ref{chi} have the
same $\chi_B=\chi_A=\chi$. Point $A$ corresponds to a city directly
connected to the seed $S$, with a flux of $w_A$ passengers per unit
time, such that $\chi_A=\ln(N\lambda/w_A)-\gamma$, while point $B$ is
obtained for a city which lies at topological distance two from the
seed, with a city $B'$ in between and such that
$\chi_B=\ln\left(\frac{N\lambda}{w_Be^\gamma}\right)+
\ln\left(\frac{N\lambda}{w_{B'}e^\gamma}\right)$. The two cases
correspond to the same value of $\chi$, but to different average arrival
times: the arrival time in $B$ is smaller than in $A$, even if
$\chi_A=\chi_B$, because relation (\ref{averaget}) neglects the travel
of additional infectious individuals between $S$ and $B'$ to compute
$\langle t_B\rangle$. Despite this systematic effect, Fig. \ref{chi}
shows that relation (\ref{averaget}) allows to define a quantity which
depends only of the populations and passenger fluxes, and such that
the average arrival time of the spreading is a monotonous function of
this quantity. In the following, we will investigate how to generalize
this quantity to more complex topologies.

\subsection{From the one-dimensional line to complex networks}

Equation~(\ref{averaget}) gives an estimate of the average arrival time in a
city $n$ connected to the seed of the spreading through links with certain
fluxes, and corresponds to a sum of the quantity $\ln(N\lambda/w)-\gamma$
along the links followed by the disease from the seed to city $n$. In
a network, 
the sum can be computed along any of the paths linking the seed
$s$ to any other city $j$. Since these different paths correspond a priori to
different fluxes and to different arrival times, a natural approach 
consists in approximating 
the average arrival time in $j$, starting from a given seed $s$, 
by the minimum of
(\ref{averaget}) over all possible paths:
\begin{equation}
\lambda \langle t_j\rangle \approx
\chi\left(j|s\right)
\equiv \min_{\{P_{s,j}\}}\sum_{\left(k,l\right) \in
P_{s,j}}\left[\ln\left(\frac{N_k\lambda}{w_{kl}}\right)-\gamma\right]
\label{defchi}
\end{equation}
where $s$ is the seed, $\{P_{s, j}\}$ is the set of all possible paths
connecting $s$ to $j$, and the sum is performed over the links
$\left(k,l\right)$ along each path. In other terms, we have
introduced a new weight on each oriented link $\left(k,l\right)$ of
the network: $\ln\left(\frac{N_k\lambda}{w_{kl}}\right)-\gamma$. The
quantity $\chi\left(j|s\right)$ is then the weighted distance between
the seed $s$ and the node $j$ on this non-symmetric network
(such a quantity can easily be computed for any topology using the
Dijkstra algorithm \citep{Dijkstra:1959}).  Note
that since the weights are real-valued, it is highly improbable that
two different paths with the same sum of weights exist, so that
equation (\ref{defchi}) selects a unique path between $s$ and
$j$. 

\begin{figure}[p]
\begin{center}
\includegraphics*[width=12.5cm]{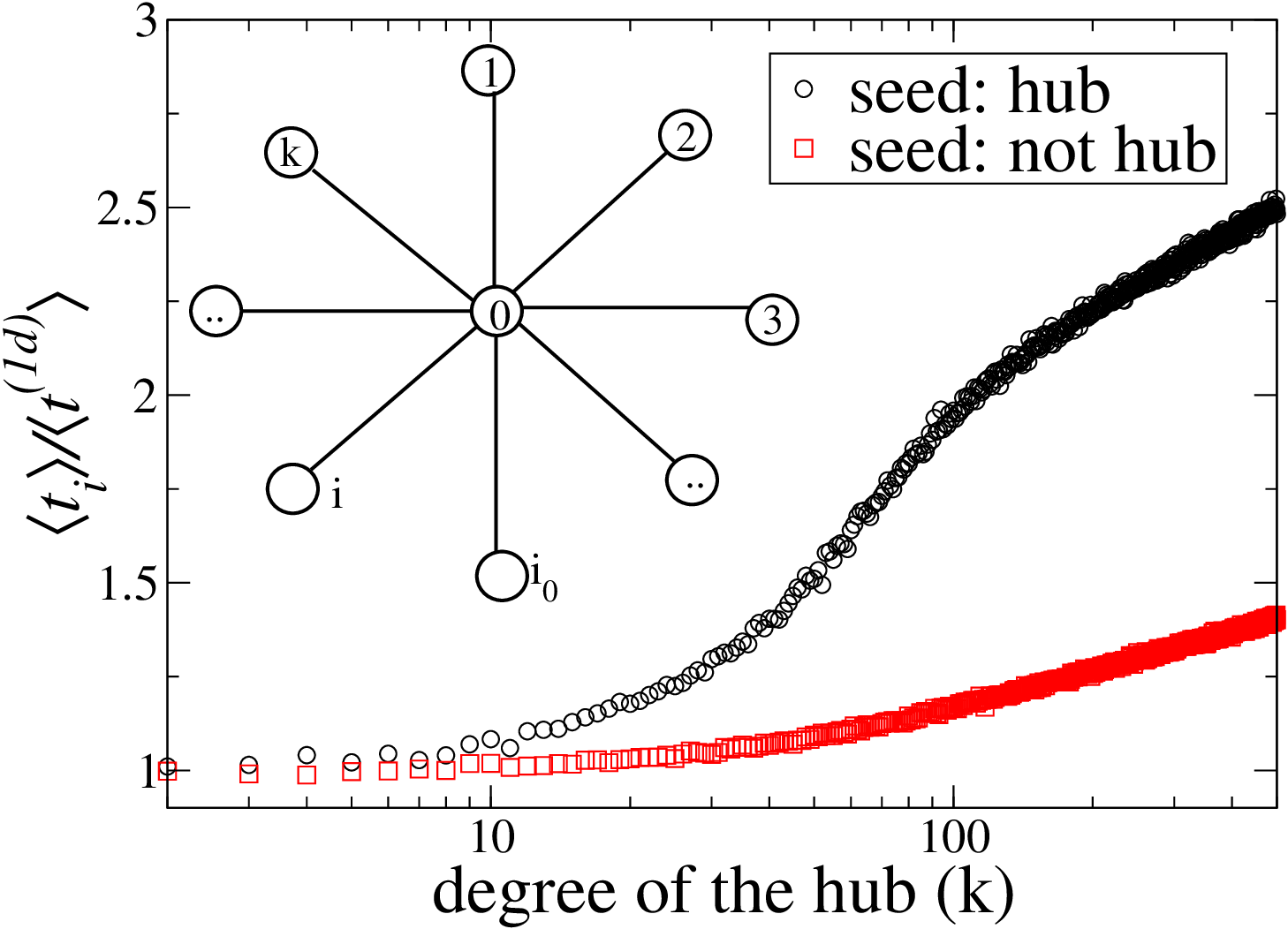}
\end{center}
\caption{ \small Ratio $\langle t_i\rangle 
/\langle t_i^{(1d)}\rangle$ of the average
   arrival time in a fixed peripheral city $i$ of the network
   shown in the inset and the average arrival time in the $1d$ case, vs
   the degree of the hub, $k$. Averages are done over $1000$
   realizations of the spreading, with parameter
   $\frac{w}{N\lambda}=10^{-2}$. Circles: the seed is the
   hub; squares: the seed is one of the peripheral
   cities. }\vspace*{0.5cm}
\label{degree}
\end{figure}

Differences between propagation on a one-dimensional line and on a
network can be expected from the two following important topological
differences between these structures: (i) paths go through nodes with
degree larger than $2$ and (ii) there are usually more than one path
between two points.  We first focus on the effect of the intermediate
nodes with possibly large degrees through which the disease propagates. To
this aim, we consider the simple topology consisting of a node $0$
connected to $k$ neighbors (as shown in the inset of
Fig. \ref{degree}). The infectious individuals, when they arrive at
the central node, have multiple possible travel destinations. The
probability for each destination to become infected is thus decreased;
if the seed is in $0$ for instance, the initial infectious individual
may even travel to a different peripherical node before creating an
endogenous epidemic growth in the hub. This effect can be quantified
by measuring the average arrival time of the disease in a given
peripheral city $i$ in various situations. We first consider the case
of a spreading phenomenon starting in a randomly chosen peripheral
node, $i_0\ne i$, and compare the average arrival time in $i$ with the
average arrival time in the one-dimensional situation in which only
cities $i_0$, $0$ and $i$ are present.  The ratio of these two times
is displayed in Fig. \ref{degree} (`seed: not hub') as a function of
the degree $k$ of the central node.  We also show in the same figure
the ratio of arrival times when the seed is $0$, i.e. the hub itself
("seed: hub"), and the topology is either the one of the inset, or the
simple two-cities configuration in which only $0$ and $i$ are present
and connected. In all cases, the arrival time in $i$ is increased by
the presence of other possible connections from the hub. The effect is
stronger for larger degree when the seed is the hub itself.  The
multiplicity of possible destinations therefore not only decreases the
predictability of the spread \citep{Colizza:2006a}, but also increases
the average disease arrival time in a given city.

\begin{figure}[p]
\begin{center}
\includegraphics*[width=12.5cm]{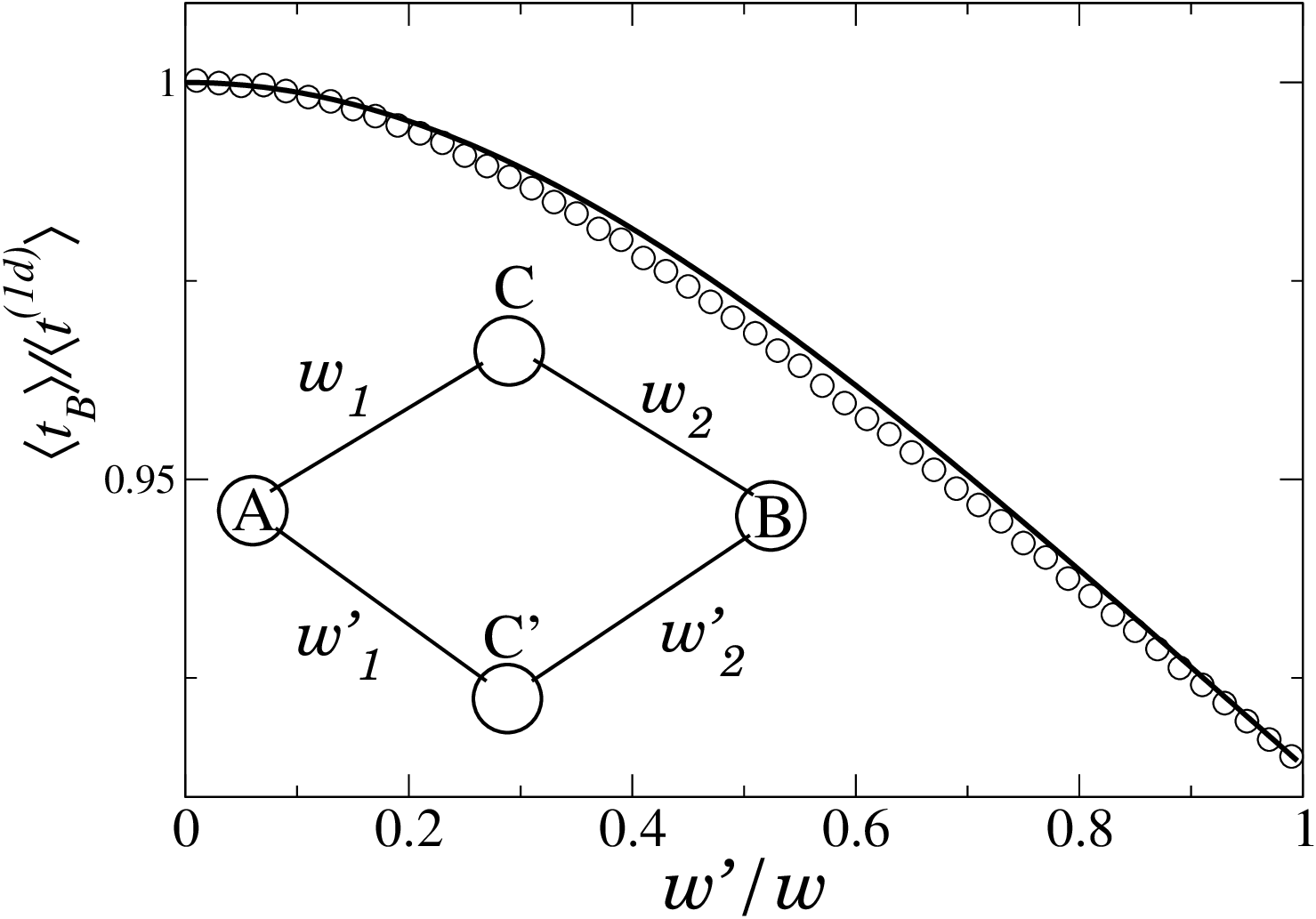}
\end{center}
\caption{ \small Ratio of the average arrival time in $B$ (the seed is
 $A$) and the average arrival time in the $1d$ case (when only the path $ACB$
 is connected), versus $w'/w$ (in the simulations $w_1=w_2=w$,
 $w'_1=w'_2=w'$, and $N=N'=N_A$). Averages are done over $1,000$ realizations of the
 spreading, with $w=10^3$ and $\frac{w}{N\lambda}=10^{-2}$. All cities have
 the same population $N=10^6$. Full line: theoretical prediction (\ref{tmp1}). }
\label{carre}\vspace*{0.5cm}
\end{figure}

Another important point distinguishes a propagation in a network from
a one dimensional line: namely, that multiple paths can often be found
between two given nodes. In fact, it has already been shown by
\cite{bart:2006} in the case in which nodes represent individuals and
not subpopulations, that the average arrival time is decreased by the
presence of multiple paths, and that not only the shortest paths
contribute to the spread between the seed and other nodes. To quantify
this effect in metapopulation models, we consider the simple topology
depicted in Fig.~\ref{carre}: a node $A$ (population $N_A$) is
connected to a node $B$ by two different paths of length two. The
first path connects $A$ with weight $w_1$ to an intermediate city $C$
of population $N$, which is in its turn connected to $B$ with weight
$w_2$.  The second path has weights $w'_1$ and $w'_2$, and the
intermediate city $C'$ has population $N'$.  We denote the probability
that the disease spread reaches the city $B$ at time $t$ by
$P\left(t\right)$ if the two paths are present, by
$P_{ACB}\left(t\right)$ if only the first path through $C$ (weights
$w_1$, $w_2$) is present, and by $P_{AC'B}\left(t\right)$ if only the
second path (through $C'$, weights $w'_1$, $w'_2$) is present. Since
the epidemics reaches $B$ from $A$ through one path or the other,
$P(t)$ can be expressed as
\begin{equation}
P\left(t\right)=\left(1-\int_0^t P_{ACB}\left(\tau\right)
d\tau\right)P_{AC'B}\left(t\right)+\left(1-\int_0^tP_{AC'B}\left(\tau\right)d\tau\right)P_{ACB}\left(t\right)
\label{min}
\end{equation}
The data shown in Fig. \ref{distrtime} suggest that we can 
assume that $P_{ACB}$ and $P_{AC'B}$ are Gumbel distributions, 
of averages
$\chi_{ACB}=\ln\left(\frac{N_A\lambda}{w_1}\right)
+\ln\left(\frac{N\lambda}{w_2}\right) -2\gamma$ and
$\chi_{AC'B}=\ln\left(\frac{N_A\lambda}{w'_1}\right)
+\ln\left(\frac{N'\lambda}{w'_2}\right) -2\gamma$, respectively. We introduce
the quantities $w_{eq}=\frac{w_1w_2e^\gamma}{N\lambda}$ and
$w'_{eq}=\frac{w'_1 w'_2e^\gamma}{N'\lambda}$. The quantity $w_{eq}$ can be
seen as the travel flux which, if $A$ and $B$ were directly connected, would
yield the same arrival time distribution than the real two links $AC$ and $CB$ with weights
$w_1$ and $w_2$:
$\chi_{ACB}=\ln\left(\frac{N_A\lambda}{w_{eq}}\right)-\gamma$.  We then have
\begin{eqnarray}
P_{ACB}\left(t\right)=\frac{w_{eq}}{N_A} 
e^{\lambda t-\frac{w_{eq}}{N_A\lambda}e^{\lambda t}}\\
P_{AC'B}\left(t\right)=\frac{w'_{eq}}{N_A} 
e^{\lambda t-\frac{w'_{eq}}{N_A\lambda}e^{\lambda t}}
\end{eqnarray}
and finally from Eq.~(\ref{min}) (using the fact that the weights are small
with respect to the quantities $N\lambda$, $N'\lambda$, $N_A\lambda$)
\begin{equation}
P\left(t\right)=\frac{w_{eq}+w'_{eq}}{N_A}\exp\left(\lambda
t-\frac{w_{eq}+w'_{eq}}{N_A\lambda}e^{\lambda t}\right) \ .
\end{equation}
We note that $P(t)$ is also a Gumbel distribution, and the average of the
arrival time in $B$, $\langle t_{mp}\rangle$ (mp=multiple paths), can be
easily calculated:
\begin{equation}
\lambda \langle t_{mp}\rangle = \chi_{mp}\equiv
\ln\left(\frac{N_A\lambda}{w_{eq}+w'_{eq}}\right)-\gamma \ ,
\label{tmp1}
\end{equation}
so that the existence of two paths results in the law
\begin{equation}
e^{-\chi_{mp}}=e^{-\chi_{ACB}}+e^{-\chi_{AC'B}} \ .
\label{multp}
\end{equation}

We have checked this prediction by numerical simulations of stochastic
spreading in the small network formed by nodes $A$, $B$, $C$ and $C'$,
in the simple case of $N_A=N=N'$ and
$w_1=w_2\equiv w$, $w'_1=w'_2\equiv w'$. 
Figure \ref{carre} displays the ratio of the average arrival time
in $B$, for a spread seeded in $A$, to the average arrival time
in $B$ when only one path (the one with weights $w$) is present. We
consider only $w' \le w$ since the situation with $w' \ge w$ is obtained
by exchanging the two paths.
As $w'\to 0$, this ratio goes to $1$, while, in the other extreme case
$w'\to w$, a decrease of the average arrival time close to  $10\%$ is
obtained. Moreover, the prediction (\ref{multp}), shown as continuous
line, is in excellent agreement with the numerical data.

\begin{figure}[p]
\begin{center}
\includegraphics*[width=12.5cm]{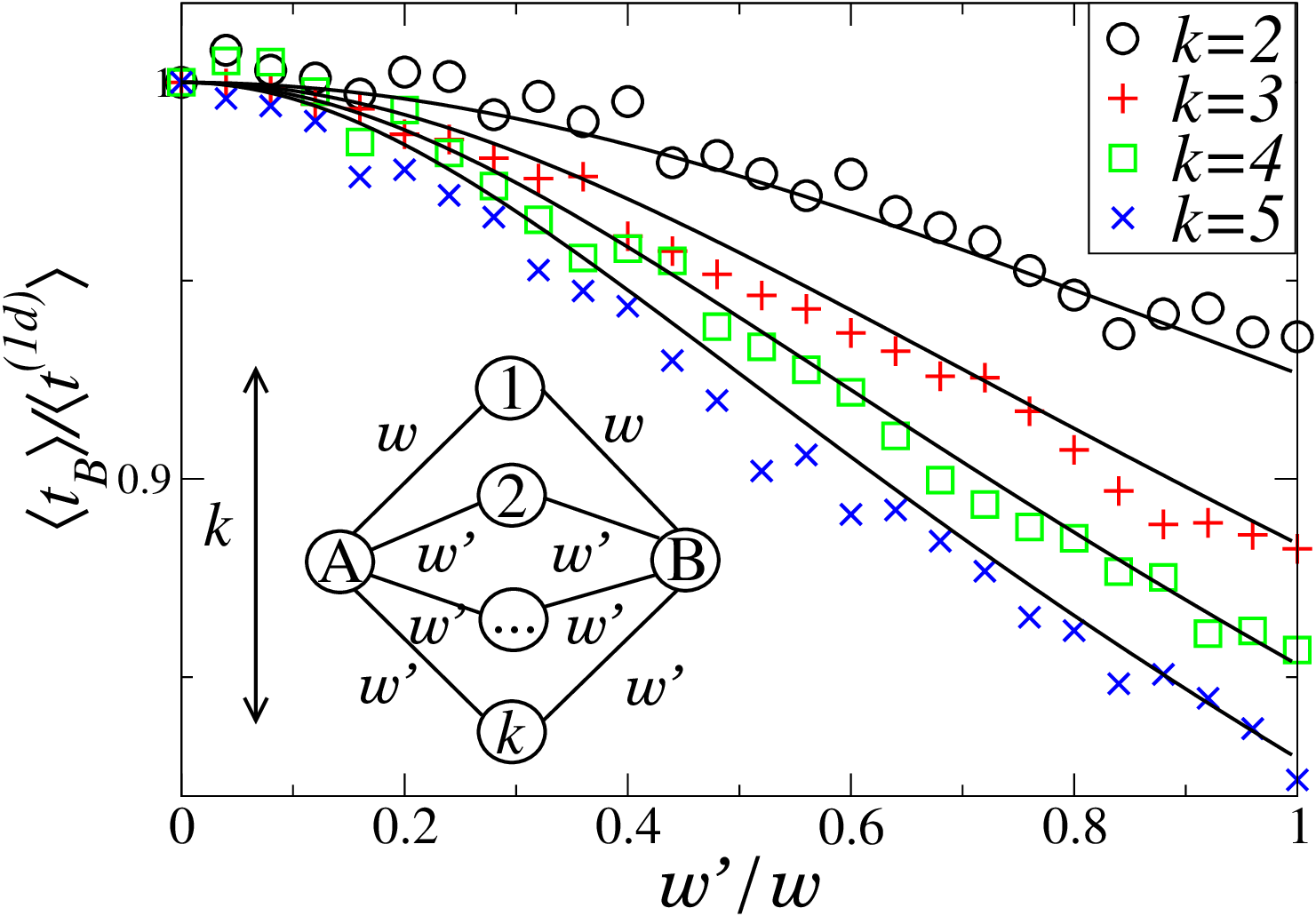}
\end{center}
\caption{ \small Ratio of the average arrival time in $B$ (the seed is
  $A$ with degree $k$) to the average arrival time in the $1d$ case
  (when only the path $A1B$ is connected), vs $w'/w$. Averages are
  done over $1,000$ realizations of the spread, with $w=10^3$ and
  $\frac{w}{N\lambda}=10^{-2}$. All cities have the same population,
  $N=10^6$.  Numerical results (symbols) are compared to the
  theoretical model given by Eq. (\ref{tmp}) (full lines) for
  different values of $k$.  } \vspace*{.5cm}
\label{cml2}
\end{figure}

Interestingly, when more than two paths are present, 
Eq. (\ref{multp}) can easily be generalized to a sum over all 
possible paths (not necessarily of length $2$): $\lambda \langle t_{mp}\rangle = \chi_{mp}$ with
\begin{equation}
e^{-\chi_{mp}}=\sum_{paths}e^{-\chi_{path}} \ ,
\label{tmp}
\end{equation}
where $\chi_{path}$ is computed on each path by Eq. (\ref{averaget}).
Figure \ref{cml2} presents a comparison of numerical results with this
analytical prediction, for various path multiplicities, showing a very good
agreement. Similar results are obtained for paths of larger lengths.  

\begin{figure}[p]
\begin{center}
\includegraphics*[width=12.5cm]{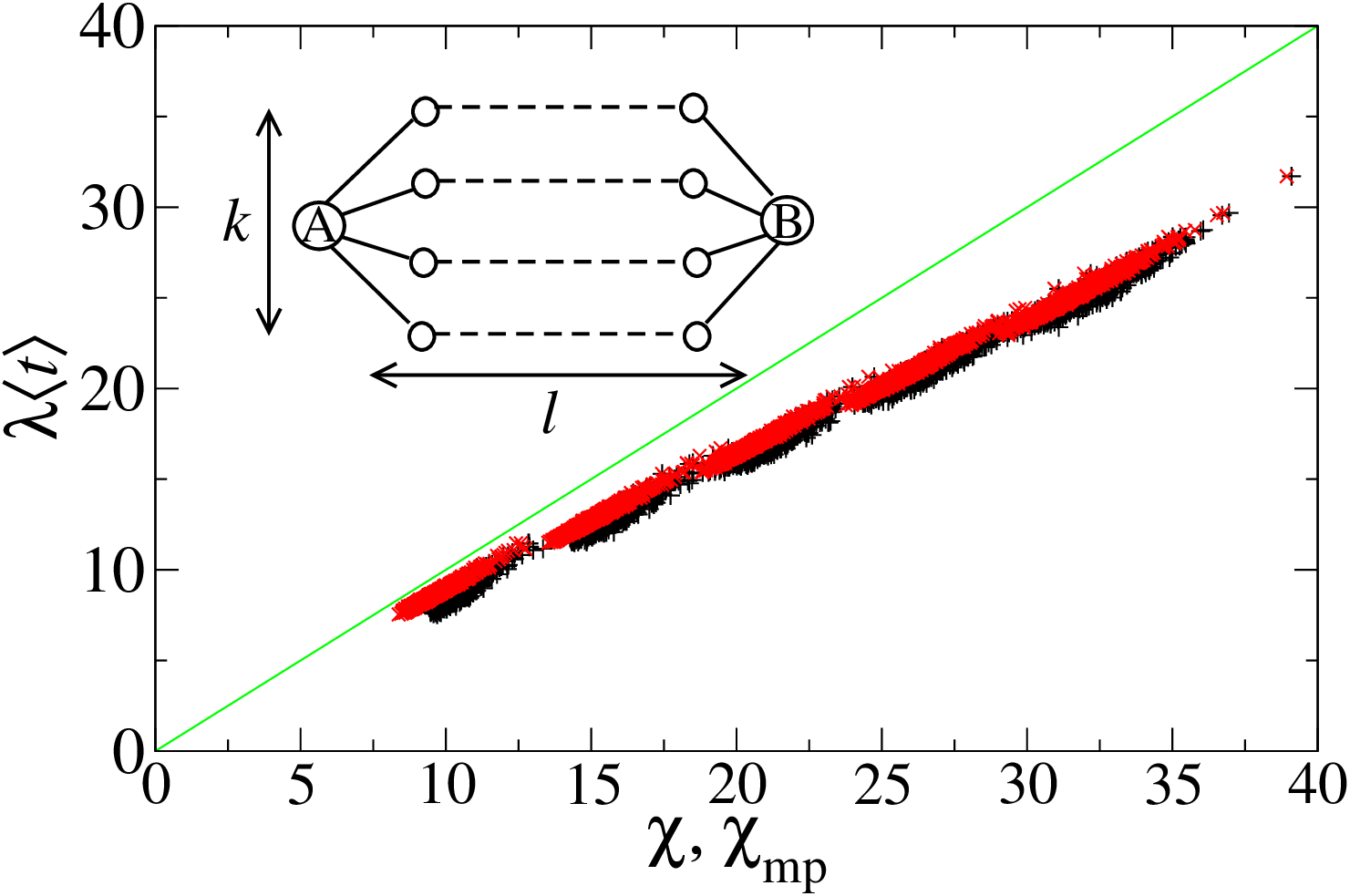}
\end{center}
\caption{ \small Correlations between the average time $\langle t \rangle$
   of arrival in $B$ and $\chi\left(B\right)$ (black), and between $\langle t
  \rangle$ and $\chi_{mp}\left(B\right)$ (red), for the network represented in
  the inset. The different points correspond to different values of $k$ (from
  $2$ to $5$), different path lengths $l$ (from $2$ to $6$) and different
  realizations of the disorder of the links ($100$ different sets of $\{w\}$
  for each structure). Averages are done over $1,000$ realizations of the
  spreading. } \vspace*{.5cm}
\label{tmod}
\end{figure}

We have also investigated numerically, on simple topologies but with random
weights, how well the average arrival time in a city is correlated with
$\chi_{mp}$ computed as in (\ref{tmp}). In particular, Fig. \ref{tmod}
displays the average arrival time in city $B$, connected to the seed $A$ of
the spreading by $k$ paths of length $l$ and random weights, as a function of
both $\chi_B$ given by Eq. (\ref{defchi}) (i.e. computed along the path that
minimizes $\chi$) and $\chi_{mp}$ which takes into account all paths. While
$\chi_{mp}$ is slightly more strongly correlated with $\langle t\rangle$ than
$\chi_B$, and closer to its actual value, the improvement is not striking. We
have also developed an algorithm to compute $\chi_{mp}$ on a complex network,
by taking into account the contributions of all paths of topological lengths
$d$ and $d+1$ (where $d$ is the length of the shortest path, in terms of
topology, between the seed and the considered city). While the algorithmic
complexity is increased (we used techniques based on the Brandes algorithm
\citep{brandes:2001}) with respect to the computation of
(\ref{defchi}), the obtained improvement was again significative but not
striking.

%---------------------------------------------------
\section{The world-wide airport network}
\label{sec:num}

We now proceed to numerical simulations of the metapopulation model on
the Worldwide airport network\footnote{We have as well considered
  artificial networks, with similar numerical results. Some
  arbitrariness in the distribution of weights and city populations is
  necessary for articial networks so that we prefer to present data obtained
  with a real-world network, in which all quantities stem from real
  data.}, which displays various levels of complexity and
heterogeneity~\citep{Barrat:2004,Amaral:2004,Colizza:2006a,Colizza:2006b}. At
the topological level, the degree distribution is broad and can be
approximated by a power-law; the links' weights
(fluxes) are also broadly distributed and span several orders of
magnitude. Finally, the world city populations are
also broadly distributed according to Zipf's law
\citep{zipf}. All these heterogeneity levels have been shown to play
relevant roles in the spread of epidemics at the worldwide level
\citep{Colizza:2006a,Colizza:2006b}. We perform the numerical simulations
of the stochastic spreading on
the network containing the $2,400$ largest airports, which takes into
account $98 \%$ of the total traffic (the WAN comprises $3,100$
airports, but we consider here only the links verifying
$\ln\left(\frac{N\lambda}{w}\right)>0$, which amounts to the removal
of the smallest airports; in particular, a certain number of nodes
have a larger traffic than inhabitants, which leads to a local
breakdown of the relation between $p_{ij}$ and the local population and 
travel flow).

\subsection{Average arrival time}

\begin{figure}[p]
\begin{center}
\includegraphics*[width=12.5cm]{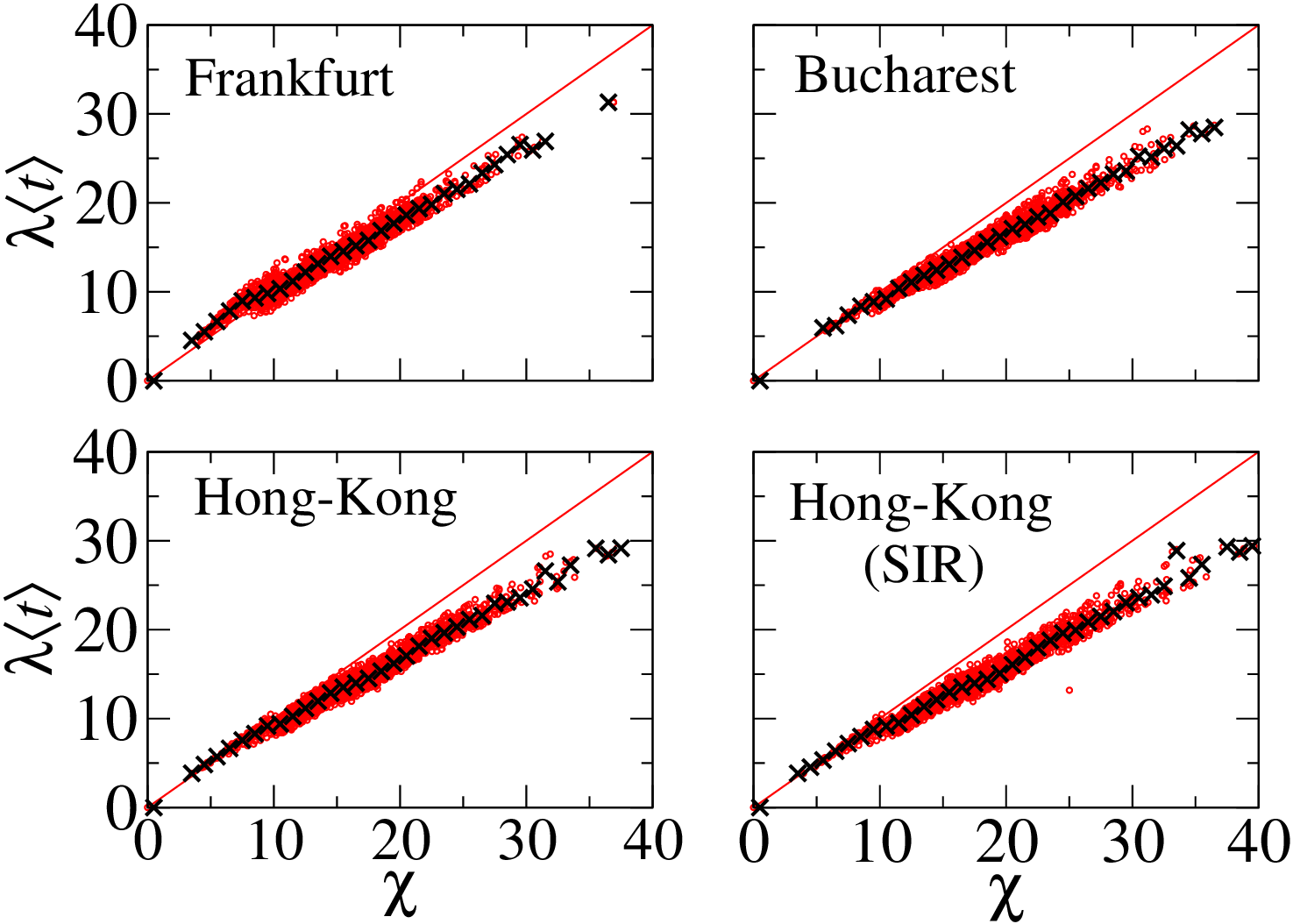}
\end{center}
\caption{ \small $\lambda \langle t\rangle$ versus $\chi$ on the WAN
     for diseases starting in different cities. Each red dot
     corresponds to a city and averages are done over $1,000$
     realizations of the spread. Crosses are an average over cities
     with the same $\chi$. The starting seeds are Frankfurt (degree
     $240$), Bucharest (degree $56$) and Hong-Kong (degree $109$).
     When the initial seed is a hub, the average arrival time is
     larger, especially in the first reached cities, due to the effect
     of the degree. We also show $\lambda \langle t\rangle$ versus $\chi$
     for a SIR model seeded in Hong-Kong. } \vspace*{.5cm}
\label{4cities}
\end{figure}

Figure \ref{4cities} displays the average arrival time in the various
cities, $\lambda\langle t_i\rangle $, as a function of $\chi_i$ as
defined by (\ref{defchi}). For each given
seed, averages are done over $1,000$ stochastic realizations of the
disease spread. The figure clearly shows that the value of $\chi$
determines the average arrival time (the two quantities are very
strongly correlated) and various cities with the same $\chi$ are
reached at the same time by the propagation.  As in the case of the
simple topologies studied above, $\chi$ in fact systematically
overestimates the correct average arrival time, but can still be
considered as a very good approximation.

Figure \ref{4cities} also highlights the effect of hubs, analyzed in
the previous section. Frankfurt is indeed the node with the largest
degree in the network, and the arrival times at a given value of
$\chi$ is larger than for other seeds with a smaller degree. This
effect remains however quite small and is essentially limited to the
first reached cities. The arrival times seem to be closer to our
Ansatz. This stems from the delay introduced by the large
degree of the initial seed, as discussed in the previous section, which
increases all the arrival times, therefore slightly shifting the data
upwards and partially compensating for the fact that $\chi$ overestimates
these times. We finally
note how very similar results are obtained for the SIR model. As
expected, the important time scale in the SIR (as well as in the SIS) is
$\lambda-\mu$: this quantity controls the endogenous growth of the
epidemic at small times. We have investigated a wide range of values
for $\lambda-\mu$, as well as for $\lambda$ in the simple SI case. The
arrival time is well determined by $\chi$ as long as the condition
$w/N\lambda<<1$ given in section 3 is satisfied.

Figure \ref{4cities} conveys the result that the 
average arrival times are not exactly given by $\chi$, but are at least 
determined to a large extent, given a seed, by this quantity.
An immediate application of this result is given by a
possible prediction of the order of
arrival of the disease in different cities. In order to compare the list of
cities ranked by the average arrival time and the same list ordered according
to $\chi$, we compute the indicator known as Kendall's $\tau$. This indicator
allows a quantitative analysis of the correlations between two rankings of $n$
objects~\citep{num_rec} and is given by
\begin{equation}
\tau=\frac{n_c-n_d}{n(n-1)/2}
\end{equation}
where $n_c$ is the number of pairs whose order does not change in the two
different lists and $n_d$ is the number of pairs whose order is inverted.
This quantity is normalized between $-1$ and $1$: $\tau=1$ corresponds to
identical rankings while $\tau=0$ is the average for two uncorrelated rankings,
and $\tau=-1$ is a perfect anticorrelation.

We present in the first column of Table \ref{tab} the values of $\tau$
for the lists of cities ordered according respectively to the average
arrival time of the spread and to the values of $\chi$, for various
seeds. The second column of the table gives the values of $\tau$ when
the lists are ordered by average arrival time and by $\chi_{mp}$. In
both cases the values are extremely high (two random permutations of a
list of size $N=2,400$ would give an indicator normally distributed
with mean $0$ and variance
$\frac{4N+10}{9N(N-1)}=2.10^{-4}$~\citep{num_rec}). In order to better
emphasize this strong correlation between the ordered list, we also
display in Table \ref{tab} the $\tau$ of Kendall between the city list
ordered by average arrival time and by different distances from the
propagation seed. As noted in the introduction, the topological
distance $d$ from the seed is not completely irrelevant, but other
weighted distances which take into account the diversity of fluxes and
populations could be expected to be more strongly correlated with the
arrival time. In particular, a first possibility consists in defining
the effective "length" of each edge as the inverse of the weight,
$\ell_{ij}=1/w_{ij}$: the disease will spread more easily on an edge
if many passengers travel across it. The corresponding weighted
distance between nodes on the network is noted $d_{1/w}$. Since the
ratios $N/w$ moreover appear naturally in the travel probabilities, we
also consider that a directed length $N_i/w_{ij}$ can be defined on
each edge, and the distance between each node and the seed can as well
be computed ($d_{N/w}$).  The results indicate significant
correlations between the average arrival times and these various
distances. The correlations are however much stronger with $\chi$ and
$\chi_{mp}$, showing that these quantities can be used with a good
confidence as an estimate of the average arrival time order. Very
similar results are obtained for SIS or SIR models, 
with respectively $\tau_{SIS}\{\langle t \rangle,\chi\}\approx
0.875$ and $\tau_{SIR}\{\langle t \rangle,\chi\}\approx 0.866$, for a
spreading process seeded in Hong-Kong.

%\end{multicols}

\begin{center}
\begin{table}[p]
\begin{center}
\begin{tabular}{l|c|c|c|c|c} 
\hline
{Seed} & $\{\langle t \rangle,\chi\}$ 
& $\{\langle t \rangle,\chi_{mp}\}$ & 
$\{\langle t \rangle,d\}$ & $\{\langle t \rangle,d_{N/w}\}$ 
& $\{\langle t \rangle,d_{1/w}\}$\\
\hline
 FRA & 0.856 & 0.917 & 0.299 & 0.311 & 0.3\\
 HKG & 0.874 & 0.934 & 0.212 & 0.275 & 0.262\\
 OTP & 0.866 & 0.938 & 0.301 & 0.304 & 0.284  \\
 SJK & 0.883 & 0.906 & 0.292 & 0.287 & 0.275
\end{tabular}
\caption{Kendall's $\tau$ for the list of cities ranked by 
arrival order and the list obtained with different indicators. Each
line corresponds to a different seed: 
 Frankfurt (FRA, degree $240$), 
Hong-Kong (HKG, degree $109$).
Bucharest (OTP, degree $56$) and 
Sao Jose dos Campos (SJK, Brazil, degree $6$).
}
\label{tab}
\end{center}
\end{table}
\end{center}

%\begin{multicols}{2}

\subsection{Arrival order for a given realization}

\begin{figure}[p]
\begin{center}
\includegraphics*[width=12.5cm]{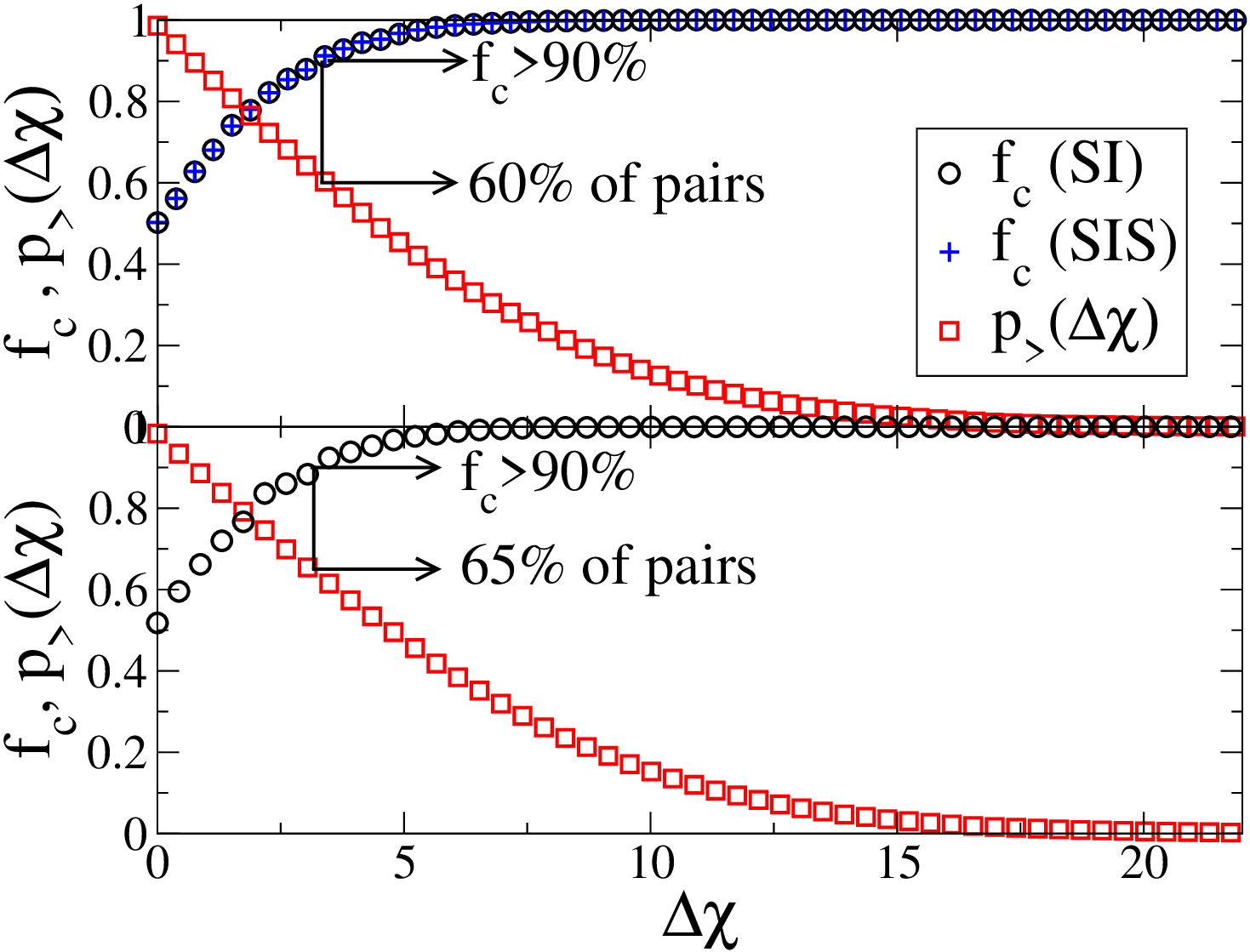}
\end{center}
\caption{ \small
  Fraction of couples of nodes correctly ranked as a function of their
  $\Delta\chi$ (circles), in each realization of the spread, and cumulative
  distribution (squares) of the values of $\Delta\chi$ (i.e., fraction of
  couples of cities $(i,j)$ with
  $\Delta\chi(i,j)=\mid\chi\left(j\right)-\chi\left(i\right)\mid >
  \Delta\chi$). Top: the seed is Hong-Kong, a node of degree $109$, and the
curves of $f_c$ for both SI (circles) and SIR (crosses) processes are shown. 
Bottom: the seed has degree $6$ (Sao Jose dos Campos, Brazil).
}\vspace*{0.5cm}
\label{proba}
\end{figure}

The previous results are valid for the average arrival times, and it
is legitimate to ask about their relevance to the case of a single
spreading event. Indeed, in the real-world there is not such a thing
as averages over different realizations. The natural extension of our
results therefore consists in checking if the quantity $\chi$ can
predict the order in which the disease will reach the various nodes
(cities) of the network in each stochastic realization of the spread.
More precisely we can compute, for each pair of nodes $(i,j)$ in the
network, and for a given infection seed, the probability that the
arrival times in $i$ and $j$ for the same realization of the spread,
$t_i$ and $t_j$, are correctly ordered by their values of $\chi$,
i.e. that they verify $(t_i-t_j)(\chi_i - \chi_j) > 0$. We use the
notation $\Delta\chi\left(i,j\right)=\mid\chi_j-\chi_i\mid$ and
Fig.~\ref{proba} displays the probability $f_c$ that a couple of nodes
with a $\chi$-difference $\Delta\chi\left(i,j\right)=\Delta\chi$ are
reached by the disease in the order predicted by their values of
$\chi$.  If $\Delta\chi\left(i,j\right)=0$, no prediction can be made
and we obtain indeed $f_c=0.5$. On the other hand, if
$\Delta\chi(i,j)$ is large ($>10$), the two nodes are reached in the
correctly predicted order in every realization of the spread.

Since not all node pairs have very different values of $\chi$ (and
thus a large value of $\Delta\chi$), we also show as in
Fig.~\ref{proba} the cumulative distribution $p_>(\Delta \chi)$ of the
number of couples of nodes with a given value of $\Delta\chi$. For
instance, for a spreading process seeded in Hong-Kong (degree $109$),
$75\%$ of the couples have $\Delta \chi > 2$, and these pairs are
correctly sorted with a probability larger than $80\%$, instead of
only $50\%$ on average if no information is available. Figure
\ref{proba} moreover shows that the precision is higher when the seed
has a small degree.  On the other hand, if the seed is Frankfurt (of
degree $240$, data not shown), the situation is slightly worsened, as
can be expected from the smaller global predictability of the spread
starting from a hub \citep{Colizza:2006a}, due to the many possible
travel destinations available for the infectious individuals. In this
case, $80\%$ of the couples have $\Delta \chi > 2$, and these pairs
are correctly sorted with a probability larger than $70\%$; the
probability $f_c$ is larger than $90\%$ for $47\%$ of the pairs.
Similar results are obtained when the endogenous growth of the disease
is described by SIS or SIR models.

\begin{figure}[p]
\begin{center}
\includegraphics*[width=12.5cm]{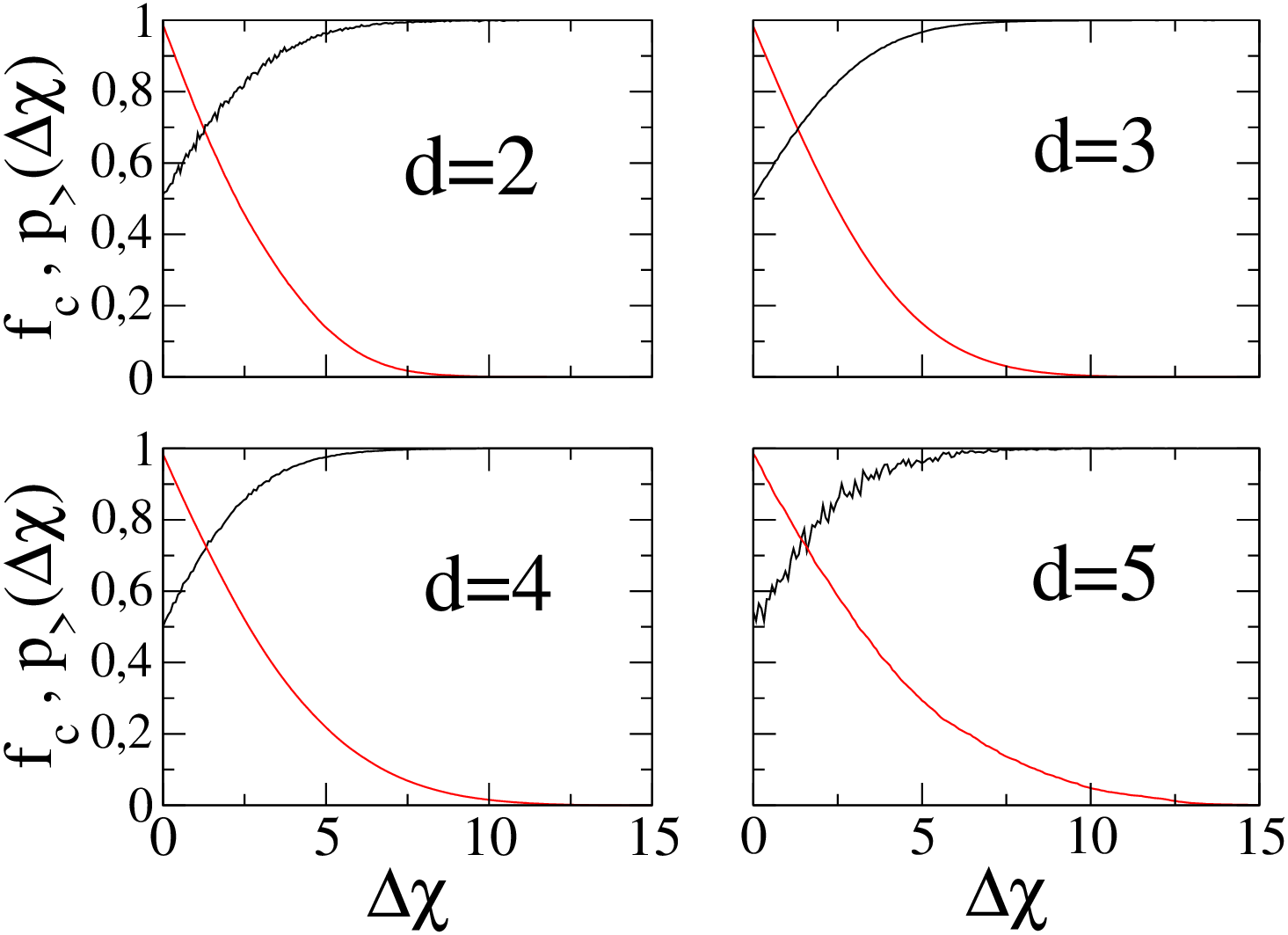}
\end{center}
\caption{ \small Same quantities as in Fig. \ref{proba}, but each graph
  concerns only couples of nodes at the same topological distance from
  the seed, $d$. From top left to bottom right : $d=2$, $d=3$, $d=4$ and
  $d=5$. The seed is Frankfurt (degree $240$).  } \vspace*{.5cm}
\label{probadis}
\end{figure}

We also present in Fig. \ref{probadis} the same quantities $f_c$ and
$p_>(\Delta\chi)$ for couples of nodes having the same topological
distance from the seed. While this topological distance is indeed too simple 
a quantity to be really useful in the prediction of arrival times, it can seem
reasonable that a node at distance $1$ from the seed will anyway be reached
before a node at distance $5$. If large values of $\Delta\chi$ were obtained
only for nodes at very different topological distances from the seed, the
prediction shown in Fig. \ref{proba} would not be particularly relevant. We
see in Fig.  \ref{probadis} that nodes at the same topological distance from
the seed can have very different values of $\chi$ and be therefore reached by
the disease in an order which can be almost certainly predicted by measuring
their values of $\chi$. For example, Fig. \ref{probadis} shows that more
than $70\%$ of the couples are well ranked with a probability larger than
$70\%$, even if the two nodes are at the same topological distance ($d=2$,
$3$, $4$ or $5$) from the seed. Results are presented for
a seed with large degree, and slightly better results are obtained
for smaller degree seeds.

\begin{figure}[p]
\begin{center}
\includegraphics*[width=12.5cm]{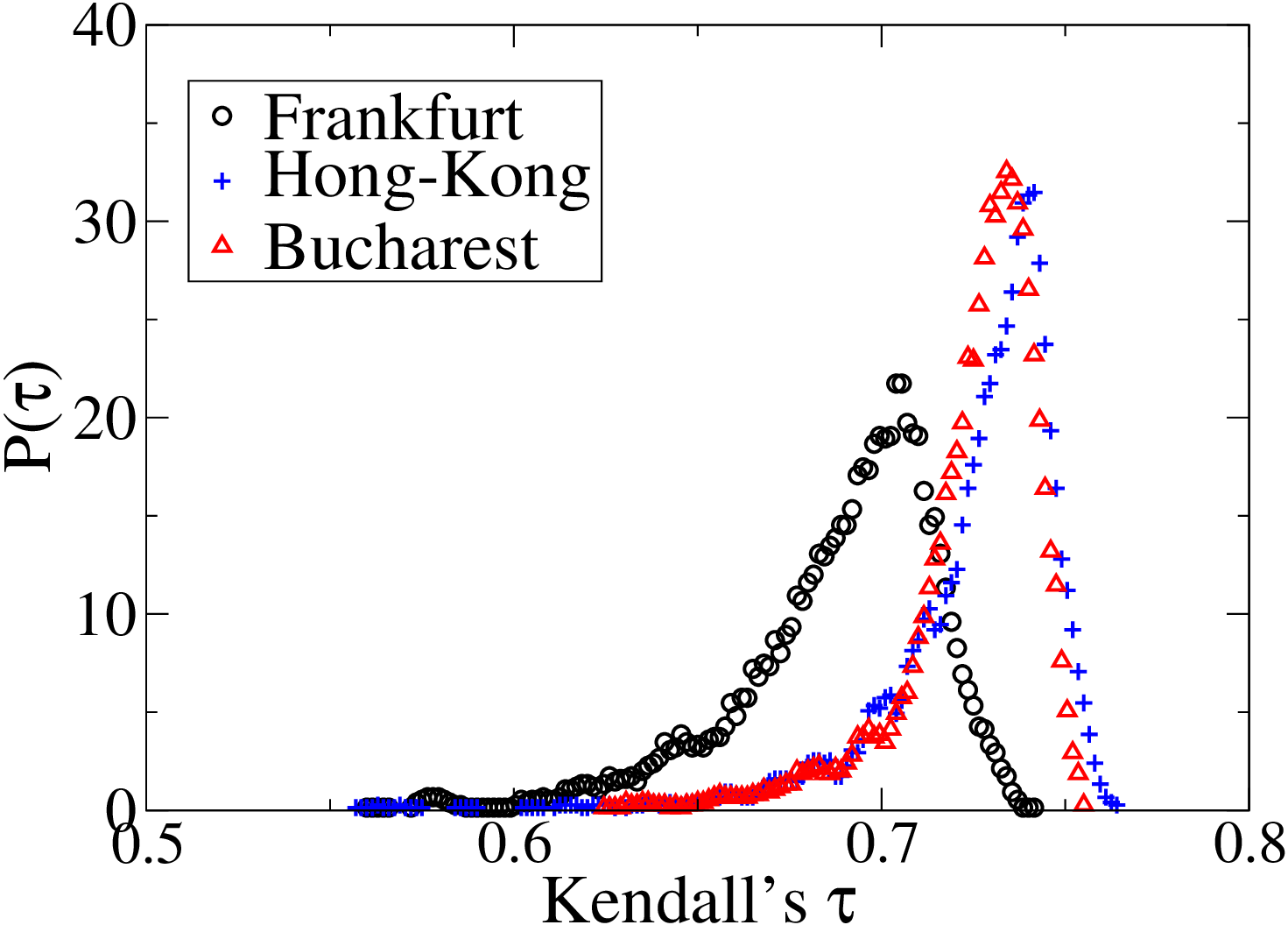}
\end{center}
\caption{ \small Histogram of Kendall's $\tau$ for 
  the list of cities sorted according either to the arrival time in a given 
 realization or to
  $\chi$. Different seeds yield different histograms.}  \vspace*{.5cm}
\label{kendall}
\end{figure}

Finally, another quantitative indication of the relevance of the
quantity $\chi$ is obtained, similarly to the previous subsection, by
comparing the list of cities ordered either by $\chi$ or by the
arrival time of the spread {\em in a given realization} (and not the
average arrival time as in the previous section). Kendall's $\tau$ is
therefore now a quantity which fluctuates from one realization to the
other (for a given seed), and the corresponding histograms are shown,
for various seeds, in Fig.  \ref{kendall}. The values obtained are
systematically larger than $0.5$, denoting a strong correlation
between the lists ordered according to the arrival time and $\chi$.
Moreover, the figure highlights how seeds with larger degrees lead to
smaller values of $\tau$.  This is in agreement with the slightly
better prediction capacities displayed in Fig. \ref{proba} when the
seed has small degree.  Interestingly, this result can also be related
to the issue of predictability studied by
\cite{Colizza:2006a,Colizza:2006b}: the predictability of a spreading
process, as measured by the similarity between two stochastic
realizations with the same initial conditions, has indeed been shown
to be larger when the seed has a small degree.

%--------------------------------------------------------
\section{Conclusion}

In this paper, we have studied a metapopulation model for the spread
of epidemics on a large scale in which subpopulations (cities) are
linked by fluxes of passengers. Such models are particularly useful
for the analysis of epidemics which propagate worldwide along the
airline connections
\citep{Longini:1985,Hufnagel:2004,Colizza:2006a,Colizza:2006b,Colizza:2007a}.
They couple endogenous evolutions of epidemics inside each
subpopulation with diffusion along the network of connections, which
opens interesting new perspectives \citep{Colizza:2007b}. In this
study, we have focused on the issue of arrival times of epidemics, as
a function of the seed and of the network's characteristics. We have
proposed a quantity easily computable on any network which depends
only on the links weights and nodes populations, and which accounts
for the average arrival time of the spread in each city. This quantity
allows to sort the various subpopulations according to the order of
arrival of each single realization of the spread with a very good
accuracy.  We note that this quantity is given by a certain weighted
distance computed on a directed weighted graph, and in particular that
this distance selects a shortest weighted path between the seed and
the various nodes. This result could therefore shed some light on the
existence and properties of ``epidemic pathways'' whose relevant role
was already suggested by \cite{Colizza:2006a,Colizza:2006b} and which
certainly deserve further work. In particular, the role and relative importance
of the multiple paths should be investigated. These predictive tools
could play an important role in the set-up of containment measures and
policies. It would also be interesting to adapt the proposed quantity
to more refined or sophisticated compartmental models
\citep{Elveback:76,Watts:2005}, or to generalize it to different
scales such as the urban scale, where nodes are places like home, work
or malls~\citep{Eubank:2004}.

\section{Acknowledgments}
It is a pleasure to thank Vittoria Colizza and
  Alessandro Vespignani for discussions. A.G. and M.B. also thank the School
  of Informatics, Indiana University where part of this work was performed. We
  are particularly grateful to Vittoria Colizza for sharing with us the data
  on the city populations and we also thank IATA (http://www.iata.org) for
  making their database available to us.  A.G. and A.B. are partially
  supported by the EU under contract 001907 (DELIS).

%\end{multicols}

\end{document}